\documentclass[preprint2]{aastex}
\shorttitle {NEW ASSOCIATION OF POST-T TAURI STARS}
\shortauthors{TORRES ET AL.}

\begin{document}

\title{A New Association Of Post-T Tauri Stars Near The Sun \altaffilmark {1}}
\author{Carlos A. O. Torres\altaffilmark {2}, Licio da Silva\altaffilmark {3}, Germano R. Quast\altaffilmark {2}, Ramiro de la Reza\altaffilmark {3} and Evgueni Jilinski\altaffilmark {3,4}} 

\altaffiltext {1}{Based on observations made under the Observat\'{o}rio Nacional-ESO agreement for the joint operation of the 1.52\,m ESO telescope and at the  Observat\'{o}rio do Pico dos Dias, operated by MCT/Laborat\'{o}rio Nacional de Astrof\'{\i}sica,  Brazil}

\altaffiltext {2}{Laborat\'{o}rio Nacional de Astrof\'{\i}sica/MCT, CP 21, 37504-360 Itajub\'{a}, MG, Brazil}
\altaffiltext {3}{Observat\'{o}rio Nacional/MCT, Rua General Jos\'{e} Cristino 77, 20921-030, Rio de Janeiro, Brazil}
\altaffiltext {4}{Pulkovo Observatory, St. Petersburg, Russia}  
\email{beto@lna.br, ldasilva@eso.org, germano@lna.br, delareza@on.br, jilinski@on.br
}

\begin{abstract}

Observing ROSAT sources in an area \mbox{$20\degr \times 25\degr$} centered at the high latitude (\mbox{$b = -59\degr$}) active star \objectname[]{ER\,Eri}, we found evidences for a nearby association, that we call the Horologium Association (HorA), formed by at least  10 very young stars,  some of them being bona fide Post-T\,Tauri stars. We suggest other six stars as possible members of this proposed association.
We examine several requirements that characterize  a young stellar association. 
Although no one of them, isolated, gives an undisputed prove of the existence of the HorA, all together practically create a strong evidence for it. In fact, 
the Li line intensities are between those of the older classical T\,Tauri stars and the ones of the Local Association stars.

The space velocity components, of the HorA relative to the Sun
%\begin{center}
(\mbox{U= $-9.5 \pm 1.0$}, \mbox{V = $-20.9 \pm 1.1$}, \mbox{W = $-2.1 \pm 1.9$})
%\end{center}
are not far from those of the Local Association, so that it could be one of its last episodes of star formation. 
In this region of the sky there are some hotter and non-X-ray active stars, with similar space velocities, that could be the massive members of the  HorA, among them, the nearby Be star \objectname[]{Achernar}. The maximum  of the mass distribution function of its probable members is  around \mbox{0.7 -- 0.9 $M_\sun$}.
We estimate its  distance as $\sim$60\,pc and its size as  $\sim$50\,pc. 
If spherical,  this size would be  larger than the surveyed area and  many other members could have been missed.

We also observed 3 control regions, two at northern and southern galactic latitudes and a third one in the known \objectname[]{TW\,Hya} Association (TWA), and the properties and distribution of their young stars  strengthen the reality of the HorA.  
Contrary to the TWA, the only known binaries in the  HorA are 2 very wide systems.
The HorA is  much more isolated  from clouds and older ($\sim$30\,Myr) than the TWA and could give some clues about the lifetime of the disks around  T\,Tauri stars. 
Actually, none of the proposed members is an IRAS source indicating an advanced stage of the evolution of their primitive accreting disks.  
  
\objectname[]{ER\,Eri} itself was found to be a RS\,CVn-like system.

\end{abstract}


\section{INTRODUCTION}  

Surveys of young pre-main-sequence (PMS) stars based on IRAS colors detect mainly the classical T\,Tauri
stars (CTT), with ages less than 10\,Myr, due to their important dusty accretion disks. Among them, one of the most comprehensive was the ``Pico dos Dias Survey'' (PDS)  \citep{12, 38, 45}.  The weak T\,Tauri stars (WTT) are mainly detected by the X-rays produced in their active coronae, since their
disks in general contain no more sufficient dust. Finally, there should exist even more evolved young stars, with masses smaller or comparable to the Sun and ages between about 10  to 70\,Myr - the so-called Post-T\,Tauri stars (PTT). If star formation were constant during 
the last 100\,Myr, a large number of uniformly distributed PTT should exist in the solar neighborhood. Nevertheless, 
the first searches for PTT resulted in very few objects \citep{14}. This lack of 
discoveries may indicate that star formation was not constant in time. 
However, as the disks around PTT are even more evolved than those around WTT, they should have been dissipated or agglutinated into planetesimals and they hardly  would be detected in  surveys based on the far infrared or even on H$_\alpha$ emission, as it  can be very weak in PTT. 

The detection of X-ray sources by the ROSAT All-Sky Survey (RASS)  \citep{30}   outside important star formation regions gave some hope that these are the expected population of PTT. 
Nevertheless, several considerations cast doubts about the PTT nature of those  dispersed objects \citep{3, 32}, indicating that most of them could  be 
active main sequence  (MS) stars. \citet{20} suggests that most of them are connected with the Local Association.  On the other hand, \citet{10} found that  some  of
those  objects could be PMS stars ejected from forming clouds. 
She also notes that the  WTT around the Chamaleon region appear to be formed in small short-lived cloudlets \citep{11}. In any case, this large population of X-ray sources around star forming clouds
seems to be represented by a mixture of PMS and young MS objects.

Do genuine PTT exist and where are they?  The main point of this work is that PTT can be found in physically 
characterized dispersed groups, with ages larger than the mean lifetime of 
the  original clouds. In such case they may be located far from any cloud. 

In fact, we discovered the first of this kind of 
association  when we were searching for new T\,Tauri stars among  IRAS point sources in a five degree 
radius around \objectname[]{TW\,Hya}, an already known isolated CTT at high galactic latitude \mbox{($b$ = $23\degr$)} \citep{8, 12}. Hipparcos parallaxes  show that this is a nearby  association,  the distance of \objectname[]{TW\,Hya} being \mbox{$56.4 \pm\,7$ pc} \citep{40}. 
Several research 
groups became interested in this association and other members were found [see. e.g.,  \citet{15, 19, 39, 18}]. Today nearly 13 young systems totalling 21 stars and a brown dwarf are known to belong to the \objectname[]{TW\,Hya} association (TWA) \citep{39, 36} and this 
constitutes a sufficiently large density of PMS stars in a localized region of the sky to consider it
as a real cluster. Futhermore, the stars appear to have a common origin \citep{24, 39} 
and the best age estimated for the association is  $\sim$10\,Myr \citep{35, 39}. 
   
Other groups of WTT, apparently isolated from clouds, have been detected  at larger distances, as the \objectname[]{$\eta$\,Cha} cluster at 100\,pc \citep{27} or the one in front of the translucent clouds MBM\,7 and 
MBM\,55 \citep{13}. The former appears to be more clearly
defined,  beeing  nearly 40 times more compact than the TWA and seems
to be related to the Sco\,-\,Cen\,OB association. Studies of this kind of clusters are important
to understand the beginning of dispersion of regions containing also massive stars. 
We propose in this paper  the existence of a new nearby association around  \objectname[]{ER\,Eri} (\objectname[]{PDS\,1} = \objectname[]{Hen\,1}), older and more isolated from clouds than the TWA.

\section{SEARCH STRATEGY AND OBSERVATIONS}

The high galactic latitude \mbox{($b = -59\degr$)} active star \objectname[]{ER\,Eri}, despite not being an IRAS source, 
was observed  during the PDS and classified as a WTT \citep{12, 45}. 
Distinctly from our earlier search for the TWA region, where candidates were selected by their infrared emission,  around \objectname[]{ER\,Eri} we used RASS sources.

We began with an area of \mbox{$10\degr \times 10\degr$} centered on \objectname[]{ER\,Eri}:
\begin{center}
 \mbox{01:30 $< \alpha <$ 02:50 and $-60\degr < \delta < -50\degr$} 
\end{center}
and later we enlarged it to about \mbox{$20\degr \times 25\degr$}:
\begin{center}
 \mbox{01:00 $< \alpha <$ 03:30  and  $-70\degr < \delta < -45\degr$}
\end{center} 
in order to find the limits of the possible association. 
We also chose three control areas, two of them  of \mbox{$10\degr \times 10\degr$}  where no isolated PMS star was known to exist $\it { a}$ $\it {priori}$.

The first control area was chosen in the southern galactic hemisphere:
\begin{center}
\mbox{06:20 $< \alpha <$ 07:40  and  $-65\degr < \delta < -55\degr$}
\end{center}

The second area was in the northern galactic hemisphere:
\begin{center}
 \mbox{09:40 $< \alpha <$ 10:20  and  $-15\degr < \delta < -05\degr$}
\end{center} 

And the third one was in the direction of the known TWA, measuring \mbox{$13\degr \times 14\degr$}:
\begin{center}
 \mbox{10:45 $< \alpha <$ 11:45  and  $-38\degr < \delta < -24\degr$}
\end{center}

As possible optical counterparts, we selected stars brighter than $\sim$13\,mag within the error boxes of the RASS,  
using the GSC and Tycho catalogues. We found 19, 46, 23, 33  and 23 X-ray sources with observable counterparts,  
respectively in the inner and  outer  areas around  \objectname[]{ER\,Eri}, in the northern,  in the southern  and in the TWA regions.
In the outer area  around \objectname[]{ER\,Eri}  we discarded, for observation, 10 bright stars, 9 being of  the Bright Star Catalogue. 
These stars are, in general,  well known and, in fact, none  is a good candidate for  their  spectral type and space velocities.
The only dwarf later than F is $\zeta^{1}$\,Ret, that has been studied recently by \citet{47}.  
These ten stars are included at the end of Table\,1 and we take them into account when doing  the statistics in this paper. 
We observed then 36 X-ray sources in this outer region.
Figure\,3 summarizes graphically our strategy around ER\,Eri.

In September 1997 and January 1998 we observed the candidates, taking low resolution spectra 
using the B\&C spectrograph installed at the Cassegrain focus of the 1.52\,m ESO telescope at La Silla, Chile. 
The  configuration used (grating of \mbox{1200 l\,mm$^{-1}$} and the CCD\,\#39) gives a final dispersion of \mbox{0.98 \AA\,pixel$^{-1}$}. 
In 1999, using the same telescope, we obtained high resolution spectra for the selected candidates, 
with the FEROS echelle spectrograph \citep{23}, which gives a total coverage of 4000\,\AA\, and a resolving power of 50000. 
To complete the extended area around \objectname[]{ER\,Eri} we took also some spectra 
using the coud\'{e} spectrograph (grating of \mbox{600 l\,mm$^{-1}$}; resolution of 9000; spectral coverage of 450\,\AA\, centered at 6500\,\AA, using the CCD\,\#101) 
of  the 1.60\,m telescope of the Observat\'{o}rio do Pico dos Dias  (OPD), Brazil.

\mbox{$UBV(RI)_C$} photometry for most of the selected stars  (except in the southern control region) was obtained  
using FOTRAP \citep{16} at  the 0.60\,m   telescope of the  OPD.
 
\section{RESULTS}

The main results of these surveys are given in Tables 1  to 4. 
In these tables we present the possible optical counterparts, the photometric data, our spectral types and the equivalent widths of the Li line $\lambda$\,6707.8\,\AA \, (W$_{\it Li}$). 
\placetable {t1}
\placetable {t2}
\placetable {t3}
\placetable {t4}

The Li\,I line, when present in late type stars, can provide a first estimate for the age \citep{20}, selecting possible PTT, moreover if the  H$_\alpha$ line is in emission or filled-in. In such way we consider young stars those with spectral types between G0 and K2 with   \mbox{W$_{\it Li} > 0.05$ \AA}, but for later types just the presence of the Li line will be a sufficient condition. For hotter stars the Li line is not a good discriminator for age (see section\,8)
and, in general, we will limit our discussion to  stars later than F in this paper.

In Tables 5 and 6 we present the kinematical and physical data of the possible young stars of the ER\,Eri and of the control regions, respectively. 
The data in italic mean deduced values, taking into account duplicity  or estimated ones, for  parallaxes and radial velocities, as we will see later. 
To obtain the Li abundances  presented in column\,13, we used our modified version of a LTE code kindly made available to us by Monique Spite, of Paris-Meudon Observatory.
These Li abundances (and the \mbox{vsin$\it i$}, column\,14) were determined by fitting  synthetic spectra to the observed ones.
In the computation of the synthetic spectra, all the known atomic lines in the range  $\lambda 6702$\,--\,$\lambda 6712$ were 
taken into account. As the molecular bands are not taken into account, the uncertainties are higher for  M type stars.  
The used stellar atmospheric models  are from Gustafsson and collaborators \citep{41}. 
For all stars we used \mbox{$\log g = 4.5$}, microturbulence velocity of \mbox{2 km s$^{-1}$} and  solar abundances. 
The Li abundance is not very sensitive to  gravity and the main source of errors   is the inaccuracy of the used effective temperatures (column\,11).

In the area around \objectname[]{ER\,Eri} there are 19 X-ray sources having possible  young stars as counterparts (8 in the inner area). The 4 visual binaries (ERX\,22, ERX\,33, ERX\,37 and ERX\,47) \footnote{For convenience, the stars will be designated in this paper by their entries in Tables 1 to 4.}  were counted as one entry each and we excluded in this count ER\,Eri\,=\,ERX\,24 for reasons we will see below.  We included in Table\,5  three F stars kinematically similar to the later stars, presenting  weak Li line (ERX\,19, ERX\,35 and ERX\,38). Nevertheless, they will be excluded in any statistical consideration.

There is an indication that a young stellar association may exist in the direction of the inner area, as the number of X-ray sources associated with PTT candidates relative to all selected sources with optical counterparts, as explained in section\,2, is greater than in all other regions. Actually, it is $\sim$40\% in the inner area, decreasing to $\sim$25\% in the additional region. This latter frequency is similar to the one of the southern control region, decreasing even more in the northern region ($\sim$10\%), as can be seen in Table\,8.

The known TWA region may furnish another useful comparison when searching for PTT associations far from clouds. 
The limits in right ascension and declination for this region were established somewhat larger than the other control regions to include the five known members of the association at the moment of the survey. We observed this region in the same way as the others and detected 9 possible young stars among 23 X-ray selected optical counterparts. 
Five of them    (TRX\,9, TRX\,11, TRX\,12, TRX\,18 and TRX\,20) are the already known members  \citep{8,12}, two (TRX\,16 and TRX\,21) were independently discovered  by \citet{36} and \citet{39}.

The frequency of possible young stars in the known TWA is very similar to the one we found in the inner \objectname[]{ER\,Eri} region. Is there a new young association in that part of the sky? To answer this question
we used several criteria   in order to characterize the young stars of this possible  association and to define its true members. These involve their kinematical properties,  relative 
proximity in the sky, ages, X-ray emission, the intensity of Li 
lines, the behavior of the H$_\alpha$ line and stellar rotation. In the following we will discuss these different criteria. 

\placetable {t5}
\placetable {t6}

\section{ KINEMATICAL PROPERTIES}

In Tables 5 and 6 we present in columns 2 to 5 the proper motions, radial velocities and parallaxes of the young stars. 
Proper motions and parallaxes measured by the Hipparcos satellite are  available for 11 of the 23 candidate members of the association around  \objectname[]{ER\,Eri} and for 6 stars of the control areas. 
For the other stars the proper motions are from the \mbox{TYCHO-2} catalogue \citep{43} (exceptions: for TRX\,12, we used \citet{39}; for ERX\,16 and SRX\,11, we  calculated the proper motions using AC2000 and GSC).  
Both catalogues are equally accurate \citep{43} and the errors are usually less than 4\mbox{ mas\,yr$^{-1}$}. These translate to tangential velocity errors of $\sim$\mbox{1 km s$^{-1}$} if the stars are at 50\,pc. 
As proper motions of stars ERX\,32 and ERX\,53 are not available, they will not be taken into account in  kinematical considerations. 

We obtained the radial velocities using FEROS or OPD  coud\'{e} spectra and their errors are similar to those in tangential velocities. 
We obtained more than one radial velocity for some stars: ERX\,4 (3 measurements), ERX\,16 (2), ERX\,21 (6), ERX\,22N (3),  ERX\,26 (2), ERX\,32 (2), ERX\,33N (2), ERX\,33S (2) and ERX\,45 (2). 
None of these stars presents noticeable variations and we assume also that the others are not spectroscopic binaries (SB). 
But we found that \objectname[]{ER\,Eri} (ERX\,24)  is a double line SB with an  orbital period of 5.9255\,days and the radial velocity used in Table\,5 is the $\gamma$ velocity (\mbox{9.0 km s$^{-1}$}).  
The analysis of this system showed that \objectname[]{ER\,Eri} is a RS\,CVn type binary at $\sim$330\,pc \citep{46} and, although being the bull's-eye for our search, it has nothing to do with the proposed association. 
This indicates that some of the other objects classified as ``young stars'' in this work may be found to be other kinds of active stars when examined in more detail.
For ERX\,35,  a known SB, we used the $\gamma$ velocity from \citet{33}. 
The radial velocities in Table\,6 for SRX\,8S, a faint optical companion of SRX\,8N,  and for SRX\,28, not measured by us, have been assumed to be 30\mbox{ km\,s$^{-1}$}, about the mean value for the region. Idem for SRX\,18, a double line SB, for which we have  only one spectrum.
For TRX\,10N we used the value of its brighter companion.

To estimate the  distances for the stars in the \objectname[]{ER\,Eri} region not measured by Hipparcos, we began supposing that they all are in the Zero-Age MS (ZAMS). 
With the distances thus obtained we computed the  space velocity components relative to the Sun, U, V (in the direction of galactic rotation) and W. 
These initial results are plotted, in the U-V plane, in Figure\,1a.
As can be visualized in the figure there is a systematic difference between the stars with the reliable parallaxes from Hipparcos  (full circles) and the others.   
The latter  are shifted   in the  direction towards  \mbox{(U, V) = (0, 0)}.
This distribution is highly improbable and it can be explained only if the distances are underestimated. 
There are two ways to produce this underestimation: 

{\it a)} Almost all  stars not measured by Hipparcos are binaries with similar components. 
This will  increase the luminosities and decrease the derived distances.
Even though possible, this is very unlikely as the only difference between the group with higher velocities and that with lower ones lies in the fact that the stars of the former are brighter and, therefore, measured by Hipparcos. 
This is an obvious contradiction with the hypothesis.

{\it b)} The stars not measured by Hipparcos are more luminous than predicted by ZAMS and, of course,  younger.

\placefigure {f1}
\notetoeditor{The best positions of Figures 1 and 2 would be if Figure\,1a and 1b ocupy two columns in the top of an even page, and fig 2 the left column of the next page. 
Another solution would be placing fig 1 and fig 2 on the same page, one  aside  the other, in landscape format.} 

If we assume this much more probable hypothesis (but keeping in mind that for some stars hypothesis (a) may be true) we can obtain kinematical parallaxes by adjusting them so as to minimize the moduli of the velocity vectors centered on the mean vector of the Hipparcos member stars.  
The space velocity components of  the 6 stars with measured parallaxes (and in this particular context we included the F stars to increase the statistical significance) are:
\begin{center}
(U, V, W)\,=\,($-8.8 \pm 1.3, -20.8 \pm 1.1, -1.4 \pm 1.5$)
\end{center}
These errors are compatible with the very low observational errors for these stars ($\sim$\mbox{0.3 km s$^{-1}$}), plus  an additional scattering  of $\sim$\mbox{1 km s$^{-1}$},  expected  for the original velocity dispersion during star formation. 
After obtaining a first set of kinematical parallaxes, we redefined the central vector including the stars within 2\,$\sigma$ of the distribution, but excluding now the F stars. 
We minimized again the moduli, but the changes were small and we stopped the process at this point. 
The distances and space velocity components obtained in this way are  presented in columns\,5 to 8 of Table\,5 and the new U-V diagram is shown in Figure\,1b. 
There is now a compact core formed by 12 stars, two of them being F stars, ERX\,22N and ERX\,19. 
One is a binary, ERX\,37, and for it we will take the mean values. 
The remaining 10 stars have the following mean space velocity components:
\begin{center}
(U, V, W)\,=\,($-9.5 \pm 1.0, -20.9 \pm 1.1,  -2.1 \pm 1.9$)\\
\end{center}

The rather large $\sigma$ in the W velocity is mainly due to the radial velocity of ERX\,14. 
To reduce it to the mean W velocity we would need to reduce the radial velocity by $\sim$\mbox{4 km s$^{-1}$}.
This could be achieved by a large, but not impossible, measurement error or ERX\,14 may be a single line SB. 
Without it we would have \mbox{W = $-1.6 \pm 1.2$}.

These ten stars form the ``probable members'' of an association which henceforth we will call the ``{\it Horologium Association}'' (HorA), since the majority of the members are in the constellation of the Horologium and we will try to prove that this is
a real association.

The next star that might be included in the HorA is ERX\,4, at 3.8\,$\sigma$ of the vector defined by the 10 stars (or 5.9\,$\sigma$ excluding ERX\,14) and we can consider this star, that lies in a corner of the surveyed area, a limit case of possible member. 
Only a very improbable conjugation of errors could bring its velocity near the mean of the 10 stars. 
The next one, the F star ERX\,35, will be at 10.6\,$\sigma$ (or 15.5\,$\sigma$ excluding ERX\,14). 
Even if we include  ERX\,4, this star will be at 6.8\,$\sigma$.
It is almost impossible that observational errors could bring this or any other selected star to the core of Figure\,1b.
We ``rejected'' these stars as members of the association  and  deduced their parallaxes as if they were on the ZAMS (Table\,5).  
We call ``possible members'', in addition to ERX\,4, the F stars in the core and the stars ERX\,32 and ERX\,53, with unknown proper motions but physically not rejectable. For reasons presented in section\,6, we classified the star ERX\,37N   as possible member. (The star ERX\,24, or ER\,Eri, is retained in this table as a ``rejected young star'', in spite of being older, for comparison to other possible mismatched cases).

The space velocity components obtained for the HorA are similar to those of the complex for the Local Association 
\begin{center}
(U, V, W)\,=\,($-11.6, -20.7, -10.5$)  
\end{center}
represented by the Pleiades and the $\alpha$\,Per groups, with ages between 20 and 150\,Myr \citep{29}. 
\citet{20} has previously noted such similarity between the kinematics of Li rich active late type stars and the Local Association. 
But for the HorA the scattering of space velocities  and the modulus of W are lower than the values  of the Local Association. 
In this sense, these values are distinct from any of the young Pleiades subgroups found by \citet{48}.

In columns\,6 to 8 of Table\,6 we give the space velocity components relative to the Sun for the control regions. 
To estimate parallaxes of the stars not in the Hipparcos catalogue of the northern and southern control areas and of the non-members of the TWA, we suppose they are on the ZAMS.   
For the TWA, we obtain kinematical parallaxes for the three stars with known proper motions in the same way as for the HorA, using as a mean vector the velocities of TRX\,9 and TRX\,18. 
With these five members the distance of the TWA is 45\,pc. 
Of course, a better estimation would be obtained using also the proposed members outside our surveyed area, but there is no published complete set of
data for none of them. 
Anyway, the distances obtained here are in agreement with \citet{24}. 
(TWA\,9 has a Hipparcos parallax of 19.9\,mas, near our mean distance for the TWA.) 
The space velocity components obtained from these five stars are: 
\begin{center}
(U, V, W)\,=\,($-11.7 \pm 1.0, -19.5 \pm 1.5, -4.7 \pm 0.7$)
\end{center}
They are close to those of the HorA.

For the control regions (Table\,6), there is a  concentration in the U-V plane at about ($-8$, $-28$) for five stars in the southern region. 
Do these young stars form another distinct physical association? Only SRX\,1 and SRX\,6 have trigonometric parallaxes - they seem mutually related - and we can use as a mean vector their velocities. 
But the obtained kinematical parallaxes for the stars SRX\,22, SRX\,18 and SRX\,28 would place them  {\it below} the MS.  Actually, the last two stars would fall far below the MS. Put in another way, their distances and the scattering in space velocities would increase if they were younger. 
The result for SRX\,11 is almost the same as if it were on the MS. 
Thus, the entries in Table\,6 represent the best we can do for this region. 
Therefore, this concentration has  discrepant parallaxes, a scattering in W velocities of \mbox{4 km s$^{-1}$} and an age near the ZAMS, rejecting it as a possible association.

Two possibilities arise: either the concentration in the U-V plane is produced by a statistical fluctuation, as it seems the case for the southern control region, or it is the result of a real association, as for the TWA region.
A fluctuation is not as improbable as it might appear at first glance since, all stars being young, they have space velocity components resembling those of the Local Association. 
Nevertheless, a real association must have a set of other characteristics alike and  that is the case for the HorA, as we will show below.

\placefigure {f2}

Another way to test if the found concentration is a distinct one is to confront it with the more homogeneous sample formed by the Hipparcos catalogue.  
In Figure\,2 we have plotted all stars measured by Hipparcos 
(excluding those that are probable or possible members of Table\,5), with distances smaller than 90\,pc in the same area of the sky of the HorA. 
To compute space velocity components, we  adopted for all stars an artificial radial velocity of \mbox{12 km s$^{-1}$} (close to the mean of the proposed association). 
In this way, the dispersion of space velocities is somewhat reduced. 
There is no concentration anywhere in Figure\,2, even at the position of HorA velocity components. This is what may be expected if the  HorA is real, formed by young active objects with constrained kinematics, and not a general behavior of the stars in the solar neighborhood. 
There are  six other stars near the core of the HorA velocity distribution and, as they could be its non-active members, we got them together in Table\,7. 
All have spectral types earlier than F6, as expected if they were the more massive and non X-ray emitting  stars of the  HorA. 
The hotter one is \objectname[]{Achernar} (\objectname[]{HD\,10144}), a well known Be star, already leaving the MS, whose evolutive age is similar to that of the proposed association. 
If we use the radial velocity given in SIMBAD, its space velocity components are near those of the proposed association ($-$6.1, $-$24.2, $-$6.3) (A in Fig.\,1). 
\placetable {t7}

\section{ APPEARANCE IN THE SKY}

In Figure\,3 we show a map of the selected X-ray sources in the  \mbox{$20\degr \times 25\degr$}  \objectname[]{ER\,Eri}  region. 
The sources not associated with young objects are evenly distributed. 
The young stars not belonging to the HorA are also almost evenly distributed (we  attributed the lack of stars at lower right ascensions  to their small number). 
The probable and possible members of the HorA seem more assembled to the center.  

\placefigure {f3}
\notetoeditor{please, this figure would be best visualized if in two columns.} 

It is yet difficult to be sure about the boundaries of the HorA, but towards the North there is a lack of members for at least $7\degr$. 
This northern strip may be converted into one more  control region. In this region, of the 20 observed plus 4 bright stars, less than 10\% may be young. 
(Now and  hereafter in this section, we will consider only stars later than F, binaries taken as one star.) 
The stars in this northern strip in Table\,5 are marked with asterisks.
Defining {\it a posteriori} the declination limits of the HorA  region  in the same way as we did for the  TWA region (that is, \mbox{$-66\degr < \delta < -52\degr$}), it contains 33 selected X-ray sources and 5 bright stars not observed, 16 being young stars (42\%), 13 of them  probable or possible members of the HorA. 
This is a 270 square degrees area and the density of selected X-ray sources is 0.14  stars per square degree, being 0.06 for all young stars and 0.05 for the  proposed members of the HorA (that is, possible and probable non-F stars).

Similarly, in the selected 182 square degrees TWA region, the densities of selected X-ray sources,  young stars and members of the association  are, respectively, 0.13, 0.05 and 0.04. 
In the northern strip  (174 square degrees) and in the northern and southern control regions the densities of X-ray selected sources and young stars are, respectively, 0.14 and 0.01;  0.23 and 0.02; 0.33 and 0.08. 
The density of X-ray sources in the southern region is much higher than in all other regions and this explains its higher density of young stars. 
There is no obvious explanation for such excess as the young stars of this region seems not  to be physically related. 
It is possible that this larger density  in the southern region reflects only a preferential exploration of the ecliptic poles by the ROSAT satellite.
Taking this higher X-ray density into account, the HorA is the densest region, slightly denser than  the inner TWA. 
(The statistics of the several classes of  selected X-ray sources of the observed regions - or sub-regions - are summarized in Table\,8.)

\placetable {t8}
Considering only the  members of both associations, the density of the HorA is higher than that of the TWA (0.05 to 0.038).
It should be noted that two of the stars found by \citet{39} in the TWA, that are not in our surveyed region, are very near our boundaries. 
But three others are far from the center of the TWA, two of them being $\sim$$10\degr$ from its apparent border.  
This shows how difficult it is to be  sure about the limits of the HorA.
The list of the proposed candidates of the TWA outside our surveyed area \citep{39, 36}  is presented in Table\,9  with our measured photometric data  (except for the bright star \objectname[]{HR\,4796}).

\placetable {t9}

The membership in the TWA is easier to establish due to its younger age. Even if its boundaries are not yet known, it is very improbable that it extends all over the sky as there is no star so young in any other of our surveyed areas.
For the HorA this is more difficult to settle on as the stars may be rather similar to  Local Association stars. 
Nevertheless, looking at the space motions in the control regions,  only NRX\,1 could be considered as a possible, but not probable, member of the HorA. (The members of the TWA, as noted before, have space motions not far from those of the HorA).

Using only the stars with known trigonometric parallaxes, the distance of the HorA would be \mbox{$46 \pm 5$ pc}. 
The mean distance, using also the kinematical parallaxes, is  $\sim$60\,pc. 
The distance, based on Hipparcos, is an underestimation, as it measures  only the brighter, that means nearer, members. 
ERX\,8, the faintest member directly measured by Hipparcos, and  also the reddest one, is of visual magnitude 9.
The kinematical parallaxes give a dispersed association, $\sim$50\,pc in diameter.
If we suppose that the original velocity dispersion  during star formation is  equal to the average modulus of the velocity vectors  (\mbox{ $\sim$1.8 km s$^{-1}$}), obtained as explained in section\,4,  the size of the HorA would be, again, $\sim$50\,pc after 30\,Myr. This size, at $\sim$60\,pc,  implies that the true angular extent of the HorA could be $\sim50\degr$, much larger than our observed region. 
In this case, many other young objects may be missing in our survey. 
A possible one is  \objectname[]{AB\,Dor}, whose space velocity components ($-8.5$, $-25$ and $-14.6$)  are not far from those of the HorA. 
Even though the W velocity is nearer to that of the Local Association, the ratio \mbox{$\log(F_x/F_b) = -3.0$} and the age of 30\,Myr \citep{6} are similar to those of the HorA.
A more extended survey is necessary to decide about such kind of possibility. 

The parallaxes for the stars of the TWA give a diameter of $\sim$20\,pc. 
Taking into account an age of $\sim$10\,Myr \citep{35, 39}, this would imply an original velocity dispersion similar to the one we found for the HorA. 
If spherical, this diameter results in an angular extent of $\sim25\degr$, which is the size needed to embrace all the members found by  \citet{39}.

\section{ AGES AND STELLAR MASSES}

An  approach to the evolutionary state  of the HorA can be made using a diagram of absolute magnitude M$_V$ versus (\bv). 
When calculating absolute magnitudes, no correction for interstellar extinction was applied as these stars are nearby and are not IRAS sources.
Actually, the possible member  ERX\,19, a F type star, is a very weak IRAS source, detected only in 12\,$\micron$, 0.24\,J, near the detection limit. 
Two other stars in Table\,5 are associated with weak IRAS sources,  ERX\,35 and ERX\,38, but they do not belong to the HorA. 
In all these cases the IRAS fluxes come from the stellar photospheres and from here on we will consider that there are no IRAS sources in the association.  

\placefigure {f4}
\notetoeditor{please, this figure would be best visualized if in two columns.}  

In Figure\,4 we present the  M$_V$ versus (\bv) diagram, its respective isochrones and evolutionary paths for some stellar masses calculated by \citet{34}.
The proposed members not measured by Hipparcos had their parallaxes estimated kinematically  as explained in section\,4. As almost all probable members are near the isochrone of 30\,Myr, the two stars without proper motions, ERX\,32 and ERX\,53, had their distances estimated supposing they are on this isochrone.

Although the distances of the majority of the stars of the HorA were obtained only from kinematical considerations, they have a small scattering in the evolutive diagram as almost all are on the isochrone of 30\,Myr, as expected for a young association. 
This is very unlikely to result by chance alone and we propose this as a first approximation for the age of the HorA.

Approximate stellar masses can be obtained by direct comparison with the positions of the stars in Figure\,4 with respect to the stellar evolution paths.
The mass distribution function  has a maximum around \mbox{0.7 -- 0.9\,{\it M}$_\sun$}.
There are some differences between this mass distribution  and that of the other young stars in the control regions. 
The HorA stars are  nearer and less massive objects than the other young  field stars and more massive than those of the TWA.
That is, most  of the HorA members are K type stars at distances from 40 to 80\,pc whereas most  of the young field stars are G type stars at 50 to 100\,pc  and the TWA members are late K or M type stars. 
Where are the M stars of the HorA and the young K-M  stars of the control regions? 
Probably beyond our observational limit of 13 mag. As we established this limit using GSC (that is, B magnitude $\sim$13), our limit for redder stars is  \mbox{V $\sim$ 12}. 
For \mbox{\bv = 1.0} on the ZAMS, this will represent stars beyond 100\,pc and for \mbox{\bv = 1.5}, with an age of 30\,Myr, stars beyond 30\,pc.
The few young K type field stars detected indicate that they are more Li depleted than those of the HorA. 
Depletion makes the Li line more difficult to be detected and, in fact, some stars with a dubious Li line in the B\&C spectra present  W$_{\it Li}$ values between 0.05 and \mbox{0.1\,\AA} when observed with FEROS. 
As the stars of the TWA are younger, the M stars are intrinsically brighter and less Li depleted, being then easier to detect.
We interpret these mass distributions as another indication that the age of the HorA is between that of the TWA and the young field stars. 
In any way, there seems to be a lack of G and early K  type stars in the TWA.

This may be not the only difference between the TWA and the HorA. In the TWA most of the stars are binaries \citep{39} and they are very few in the HorA. 
In fact, with the instruments used, we are unable to detect close visual binaries in the HorA - ERX\,22 and ERX\,37 being very wide. The primary of ERX\,37, lying above the 30\,Myr isochrone, is a candidate to be investigated for duplicity. 
As the other stars are on  the isochrone (except for ERX\,16, see below), any eventual secondary must have a very large luminosity difference.
We have very few radial velocity measurements to detect single line SB.
Nevertheless, at the sky position of the HorA, the radial velocity is mainly in the W direction. 
Even if, in general, we have not measured the stars more than once, the slightly larger scattering in the W velocity, obtained in section 4, shows there is only a small possibility that some of these stars are SB. The best candidate is ERX\,14. 
For the TWA we can compare the observations of this survey to those of the PDS and, at least, there is one SB - TRX\,20 (\objectname[]{PDS\,55}). 
Double line SB might be detected in only one exposure and we found five previously unknown SB (see Tables\,1, 2 and 3), but none is a proposed member of the HorA.    

Even considering this limited information we have an indication that the frequency of binaries in the HorA is less than in the TWA. The existence of some yet undetected binaries may decrease  the scattering in the evolutive diagram even more. 

\subsection{Comments on some objects}

Some of the objects in Figure\,4 deserve  detailed comments:
\paragraph{ERX\,4}  (\objectname[]{CPD\,$-$64\,120}): This star is in the lower right corner of Figure\,3 and has discordant space motions. It lies on the ZAMS and this may be another reason to exclude it as proposed member of the HorA. But, as its parallax was  determined kinematically, the accidental errors that would conspire against suitable space velocity components could also underestimate the distance. We have observed thrice for radial velocity with similar values. It presents large variations in H$_\alpha$ - from absorption to a small emission.
Thus, we keep ERX\,4 as a limit case of membership. 
\paragraph{ERX\,14}  (\objectname[]{GSC\,8047-0232}): The probable member with the large discordant W. Its position in the evolutive diagram is compatible with the supposition that it may be a single line SB. We have only one radial velocity observation of this star.

\paragraph{ERX\,16}  (\objectname[]{CD\,$-$53\,386}): It is the only probable member having a discrepant position in the evolutive diagram, and, besides, this is  the only star in Table\,5 without \mbox{TYCHO-2} proper motions. Nevertheless, to check our proper motions, we used also the PPM and the USNO-A2, without any appreciable difference.
For its absolute magnitude to be compatible with the age of 30\,Myr, the distance should be $\sim$80\,pc. 
This would result in \mbox{(U,V,W) = -6.2, -16.8, -5.1}. Even not giving a perfect fit to the kinematics or age test to belong to the HorA, it is still close enough to keep it as probable member.

\paragraph{ERX\,22}  (\objectname[]{HD\,13246}):  The stars of this system  are separated by  $52.76\arcsec$ but whereas  star ERX\,22N was not directly measured by Hipparcos, \mbox{TYCHO-2} gives very similar proper motions and  \citet{37} has not found changes in their relative position  since 1919, indicating that both stars constitute a physical pair separated by 2380\,AU. 
Therefore, the parallax of ERX\,22N was taken as being equal to that of the primary.

\paragraph{ERX\,33}  (\objectname[]{HD\,16699}):   This visual binary has been detected as such by Hipparcos,  with a separation of  $8.74\arcsec$, and our identical measured values of the radial velocities confirm it as a  physical system. 
ERX\,33S seems younger than  the HorA by its position in the evolutive diagram. 
The spectrum of the primary (ERX\,33N) has a cooler system of lines superimposed (like a late G or early K star) and \citet{44}  explain it by a close visual binary. (Actually, the USNO-A2 has two stars - the same? - in the field, of 11.5 mag., but they hardly  could be the faint component.)   
\citet{44} measured also the  \mbox{vsin$\it i$} of this system and found almost the reverse of our values. 
One possibility is that ERX\,33S is a double line SB too, not resolved in both works. 
This could explain the positions in the evolutive diagram not in good agreement for coeval stars and, in this case, more compatible with  an age similar to that of the HorA.  
Anyway, these stars should not be considered  members of the HorA due to their space velocities.

\paragraph{ERX\,37}  (\objectname[]{CD\,$-$53\,544}):  This system had already been proposed by  \citet{21} as an  isolated pair of PTT. 
The separation is about $22\arcsec$ and both stars are X-ray sources,  the hotter one (ERX\,37S) having a stronger flux, compatible with its larger \mbox{vsin$\it i$}  (\mbox{80 km s$^{-1}$}) - one of the highest for this association. 
Our radial velocities are similar   for both stars, as also found by \citet{21},  indicating  that they could form a physical pair.   
The main problem is the large difference in Li abundance between the components. One possible  explanation is that ERX\,37N may be a field dMe star and the stars would not form a physical pair, only ERX\,37S being member of the HorA. 
Nevertheless, if ERX\,37N has an age of 30\,Myrs, it would be at the same distance as the kinematical one obtained for ERX\,37S, and this can hardly be accepted as being due to chance. 
Another possibility, adopted by \citet{21}, takes into account  a proposed dependence of Li depletion with rotation. 
Our measured low Li abundance ($-$1.2) for the slow rotator ERX\,37N is in the direction of such proposal. 
Cautiously, we will  consider ERX\,37N as only a possible member of this association.
ERX\,37S is above the 30\,Myr isochrone and this discrepant position  could be an indication for duplicity.

\paragraph{ERX\,32 and ERX\,53}  (\objectname[]{GSC\,8056-0432} and  \objectname[]{GSC\,8499-0304}): The distances for these stars are estimated supposing they are on the isochrone of 30\,Myr, as they have no measured proper motions.
But they appear to be members of the HorA for their large X-ray fluxes and large  W$_{\it Li}$, as we will see in the following sections. 
Actually, the W$_{\it Li}$  for ERX\,32 seems very high compared to the depletion expected for a M3 star, but its \mbox{vsin$\it i$} is also high for this spectral type.

\section{  X-RAY FLUXES AND ROTATION}

In column\,10 of Tables\,5 and 6 we present the  ratios of X-ray fluxes to total luminosities,  following the  procedure of \citet{19} to obtain the actual X-ray fluxes.  
In double systems we suppose that the X-ray fluxes are equally contributed by both components (except for ERX\,37, for which we used the individual X-ray fluxes from \citet{21}).  
The values for the probable and possible members are high  and compatible with those expected for PMS stars \citep{7} - except for  the F stars ERX\,19 and ERX\,22S  -   and they are near the empirical saturation level for the Pleiades age \citep{24}. 
In fact, the mean ratio $\log(F_x/F_b)$ for the  probable members  \mbox{($-3.2 \pm 0.2$)} is compatible with the evolutionary age \citep{24} and similar to that we  obtained for the TWA  \mbox{($-3.0 \pm 0.3$)}. 
Only half of the young field stars have such high values.

Among the probable or possible members of the HorA, there are five stars with  \mbox{vsin$\it i$ $>$ 30 km s$^{-1}$} and  four  
among young field stars:  one in the northern strip, ERX\,26, two in the border of the HorA region, ERX\,50 and ERX\,54,  and one SB, SRX\,18. 
There is also one star in the TWA, a SB too, TRX\,20.

The $\log(F_x/F_b)$ values for the probable and possible members of the HorA, excluding F stars, are weakly correlated to  \mbox{vsin$\it i$}  \mbox{(R = 0.64)}. 
Although this is expected if rotation is one of the main physical causes of the X-ray emission in late type stars, even this weak correlation is not easily obtained since  differences in  stellar inclinations and evolutionary stages  smear out the correlations for non-coeval stars. 
In fact, there is nothing like that for the young field stars. 
As almost all measured stars  of the TWA present small \mbox{vsin$\it i$} it is impossible to draw any conclusion for this association.

The rotation of \objectname[]{ER\,Eri} shows that, if this star is synchronised, \mbox{Rsin$\it i$ = 4.4 R$_\sun$}, which is incompatible with a WTT. 
Actually we found that this star is a RS\,CVn type system \citep {46} at 330\,pc  and it was only by a very good luck that we found the HorA taking it as the bull's eye of our search.

\section{ LITHIUM AND H$_\alpha$ LINES}

Because PTT are in an intermediate evolutionary stage between CTT or young WTT and active MS stars, genuine PTT are expected to have a peculiar H$_\alpha$ 
behavior \citep{31}. 
Actually, the  W$_{\it H\alpha}$ of the members of the HorA behave in a monotonic mode with respect to temperature, with  fill-in at  \mbox{\bv $\sim$ 0.9} and in emission for later type stars. 
For the young field stars there is more scattering, but the different distributions of stellar temperatures between both samples makes a detailed comparison difficult.

\placefigure {f5}
\notetoeditor{please, this figure would be best visualized if in two columns.} 

One of the best confirmations of the PTT nature of the stars, whose membership to the HorA was established by  their space motions, comes from  the Li lines. 
\citet{28}, comparing the behavior of the  W$_{\it Li}$  with temperature for a sample of CTT and stars of some young clusters, including the Pleiades, finds a gap (the PTT-gap) for stars cooler than 5250\,K, between the high  W$_{\it Li}$  formed by CTT and the lower  W$_{\it Li}$ formed by young MS stars belonging to the  Local Association. 
A similar conclusion is obtained by \citet{20}. 
The Li abundance is not a good criterium to discriminate PTT above the cutoff at \mbox{T$_{\it eff}$ = 5250 K}, as there is almost no Li depletion in hotter stars.

The  W$_{\it Li}$  of the stars of the HorA are mainly in the lower edge of the PTT-gap (Fig.\,5). 
The figure shows that the HorA is younger than the youngest open clusters of the Local Association plotted by \citet{28} and  \citet{20}.
The two stars of the HorA with  W$_{\it Li}$ similar to the ones of the young field stars have the smallest  \mbox{vsin$\it i$}.

Ten young field stars are cooler than the cutoff in temperature, but only two of them may be younger than the Local Association: SRX\,8S and TRX\,10N. 
The latter seems a physical binary whose primary has a  W$_{\it Li}$ normal for a young star of its temperature, whereas the weak value of the secondary was obtained from a low resolution spectrum and is of low quality, making it a poor candidate for PTT. 
SRX\,8S, the optical companion of the brighter SRX\,8N,  had its W$_{\it Li}$  measured also on a  low resolution spectrum. 
Its W$_{\it H\alpha}$ indicates some filling-in, but as it is very near the temperature cutoff, it is not a good PTT candidate too.

The stars of the TWA are on the upper edge of the PTT-gap, and some of its members are similar to CTT or WTT. 
The gap between the TWA and the HorA stars in Figure\,5 indicates the amount of Li depletion in $\sim$20\,Myr.

\section {DISCUSSION}

In view of all the above arguments, we can conclude that there is, in the direction of the star \objectname[]{ER\,Eri}, an association of very young stars, younger than the majority of the groups of the Local Association and older than the TWA, characterized by the space motions and with properties that we have not found in other parts of the sky.  We named it the ``Horologium Association''. Its isolation from clouds is quite remarkable. 
Even if there are some dispersed infrared cirrus at a few degrees from its center \citep{9}, the nearest clouds are  at $25\degr$ from it.
Moreover, these  are  CO translucent clouds, where there is no evidence for star formation \citep{26, 4}. 
In any case, considering the proposed age for this association  ($\sim$30\,Myr), it is hopeless trying to find its birth place.

It is also interesting to note that contrary to the TWA, which is formed by a mixture of a few CTT as \objectname[]{TW\,Hya} itself and a larger number of
PTT, the HorA is formed only by  PTT where none of them is an IRAS source, indicating the absence of dusty accretion disks. 
This suggests that, at their advanced evolutionary stage, the circumstellar disks have been dissipated or agglomerated into planetesimals and, therefore, the HorA can give an upper limit for the lifetime of the disks.    
\citet{18} discussed the problem raised by the existence of some isolated CTT having active disks  in the TWA, as  \objectname[]{TW\,Hya} itself. 
There is also the case of the visual binary \objectname[]{PDS\,50} in the same cluster,  where
the component B could be evolving from a CTT to WTT stage \citep{17}. 
We  suppose that the TWA will evolve into the HorA stage and that the disks around their
CTT will loose the IRAS properties in a time scale of less  than 30\,Myr.
Considering that the majority of objects in the TWA are WTT, the few CTT would be in their last stages of activity. 
The transformation of the disks would consist in converting their dust into pre-planetary material. 
Quite recently \citet{42} have detected  a planet of 2.2\,Jupiter masses in an Earth-like orbit (320\,days)  around the star \objectname[]{$\iota\,$Hor} (ERX\,38), one of the rejected members in Table\,5. 
As this star seems not to belong to the HorA,  we could not use the age of this association to get a better idea of the time scale  of the formation of a planet around  a young star. 
On the other hand, \citet{2} propose that the protoplanetary disk of \objectname[]{$\beta$\,Pic} has only 20\,Myr, less than the age of the HorA. 
Probably there is  no universal time scale for the apparition of planets, that should depend on the properties of the accreting disk.

\section{CONCLUSIONS}

Exploring a region of about \mbox{$20\degr \times 25\degr$}  around the high galactic latitude \mbox{($b = -59\degr$)}   active star  \objectname[]{ER\,Eri},
previously classified as a WTT  \citep{12, 45}, we found evidences of a new  association, the ``Horologium Association'', formed mainly by bona fide low mass \mbox{($\sim$0.8 {\it M}$_\sun$)} PTT. 
Its probable and possible members are marked with asterisks in Table\,1 and are discriminated in Table\,5. Possible hot members are presented in Table\,7.
\objectname[]{ER\,Eri} itself was found to be a  background RS\,CVn-like  system.
Since we found no low mass PTT in the two control areas, as can be seen in Figure\,5,  we believe that HorA-like stellar groups are not numerous in the solar vicinity, indicating a non constant rate of star formation in the last 100\,Myr. 
This new association is presently represented by at least 10 members (Table\,5), having an age of $\sim$30\,Myr and is older than the isolated TWA, with an age of approximately 10\,Myr.  
Until now there are no detected binaries in the HorA (stars ERX\,22N and ERX\,22S have a separation of 2380\,AU and should be considered as common proper motion stars).  
If confirmed, this marked difference with the TWA must be due to different intrinsic conditions of star formation of these two associations. 
The HorA is even more isolated from clouds than the TWA and, in any case, we can expect that its original cloud could have dissipated in less than 30\,Myr.

The distances of the stars in the HorA cover an interval  from nearly 40 to  about 90\,pc (at a mean distance of $\sim$60\,pc) giving a diameter of $\sim$50\,pc, compatible with the size produced after 30\,Myr  by  an initial velocity dispersion of \mbox{$\sim$1.8 km s$^{-1}$}.  
If this size  is the same in angular extent, it surpasses the surveyed region and  could contain many other  
members, such as the interesting young star \objectname[]{AB\,Dor}. 
In the surveyed region there may be hotter members, that are not X-ray sources,  (for example, the nearby Be star \objectname[]{Achernar}) having similar ages, distances and space motions.  
If this is the case, the HorA could be the remnant of an old OB association, but, due to the kinematical errors, there is little hope to localize its birth place.

The space velocity components of the HorA are near but distinct from those of the Local Association.
In fact, in all surveyed regions we found young field stars, possible members of the Local Association, with a compatible Li depletion. 
But, for its well defined kinematics,  physical properties, Li abundances and restricted  location in the sky, the HorA  can be distinguished from these stars, resembling more the general behavior of the TWA. 

Considering the observational limitations, as that of the magnitude limit of the Hipparcos measurements, we  are aware of the
challenge that represents the detection of a coeval moving group of stars with an age around 30\,Myrs. 
To arrive at this, we examine several requirements that characterize  a young stellar association. 
If no one of them, isolated, is completely sufficient, none of them are mutually contradictory and all together practically create the necessary condition for the establishment of the HorA as a real nearby association. 
Following \citet{5} and considering that large moving groups, more commonly known as Eggen's superclusters, are not real clusters 
but are  formed by a chance coincidence of smaller coeval streams, we can conceive the HorA as one small, coeval structure in the Pleiades supercluster.
But its overall characteristics are distinct from those of the young Pleiades subgroups found by \citet{48}.
In any case, the HorA may be one of the  last episodes of star formation of the Local Association  and  could be useful to understand better its fine structure.

\acknowledgments

We thank L. Siess for providing the theoretical evolutive diagram used in this work. Drs. S. Frinck,
B. Reipurth, G. Cutispoto, A. Andrei and N. Drake for instructive discussions. L. da Silva thanks ESO for the data
reduction facilities. 
The  suggestions  and careful analysis  of an anonymous 
referee contributed substantially in the improvement of this paper.
We thank also the Centre de Donn\'{e}es Astronomiques de Strasbourg (CDS) and NASA for the use of their electronic facilities, specially SIMBAD and  ADS. This work was partially supported by the
CNPq,
grants to L da Silva and R. de la Reza, processes 200580/97 and 301375/86,
respectively.

\clearpage

\figcaption [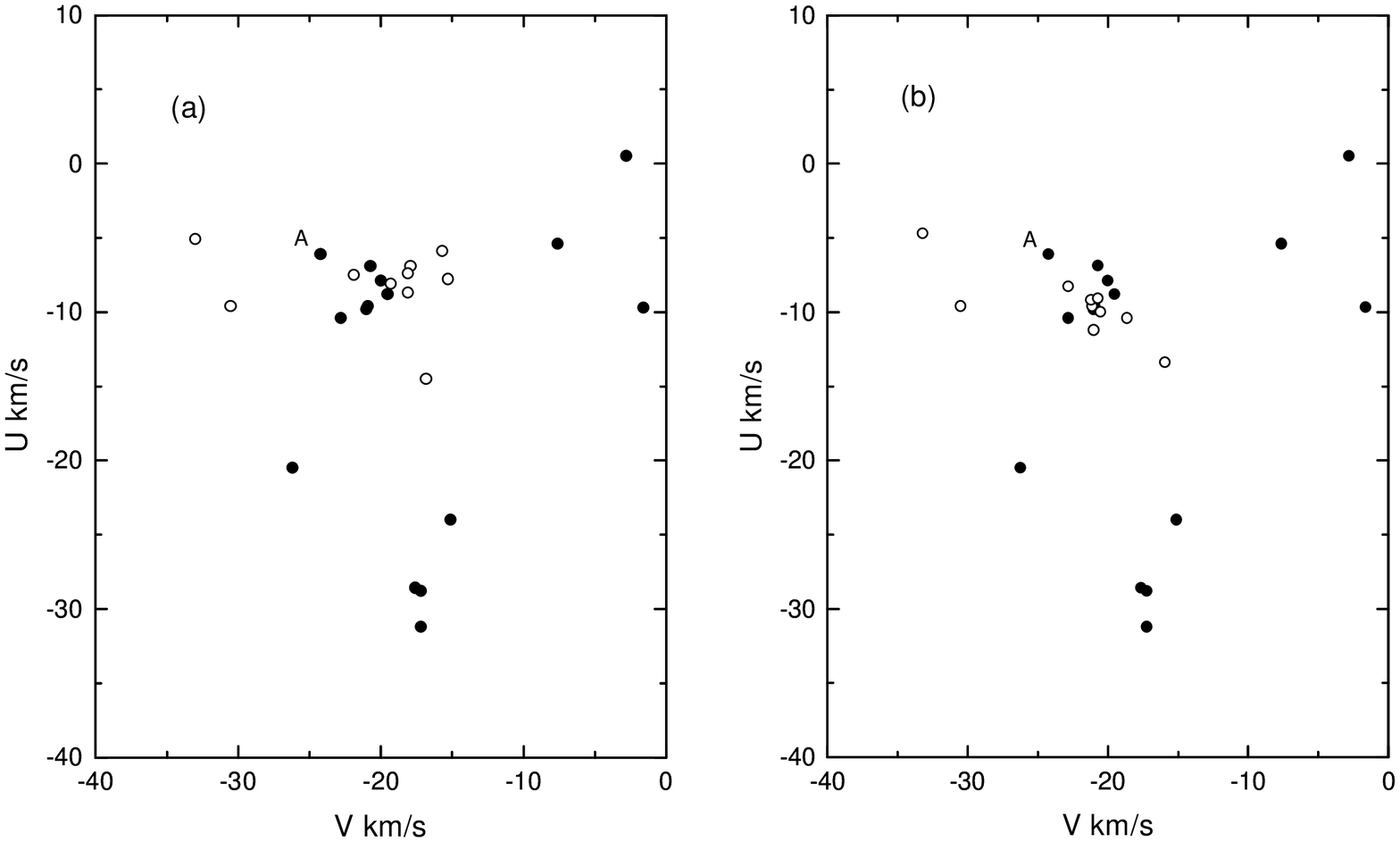] {(U, V) space velocity components for stars in the ER\,Eri region (Table\,1) with Hipparcos and estimated parallaxes for candidate stars. 
Some high velocity stars are outside the figure. 
Open circles are stars with estimated parallaxes: 
(a) supposing  they were on the ZAMS, showing a systematic shift relative to the central core; 
(b) kinematically, obtained by minimizing the velocity scattering.
The stars "move" grossly in the direction away from the position (0, 0). 
The central concentration defines the Horologium Association. 
Achernar (A) is plotted for comparison.    \label{f1}} 

\figcaption [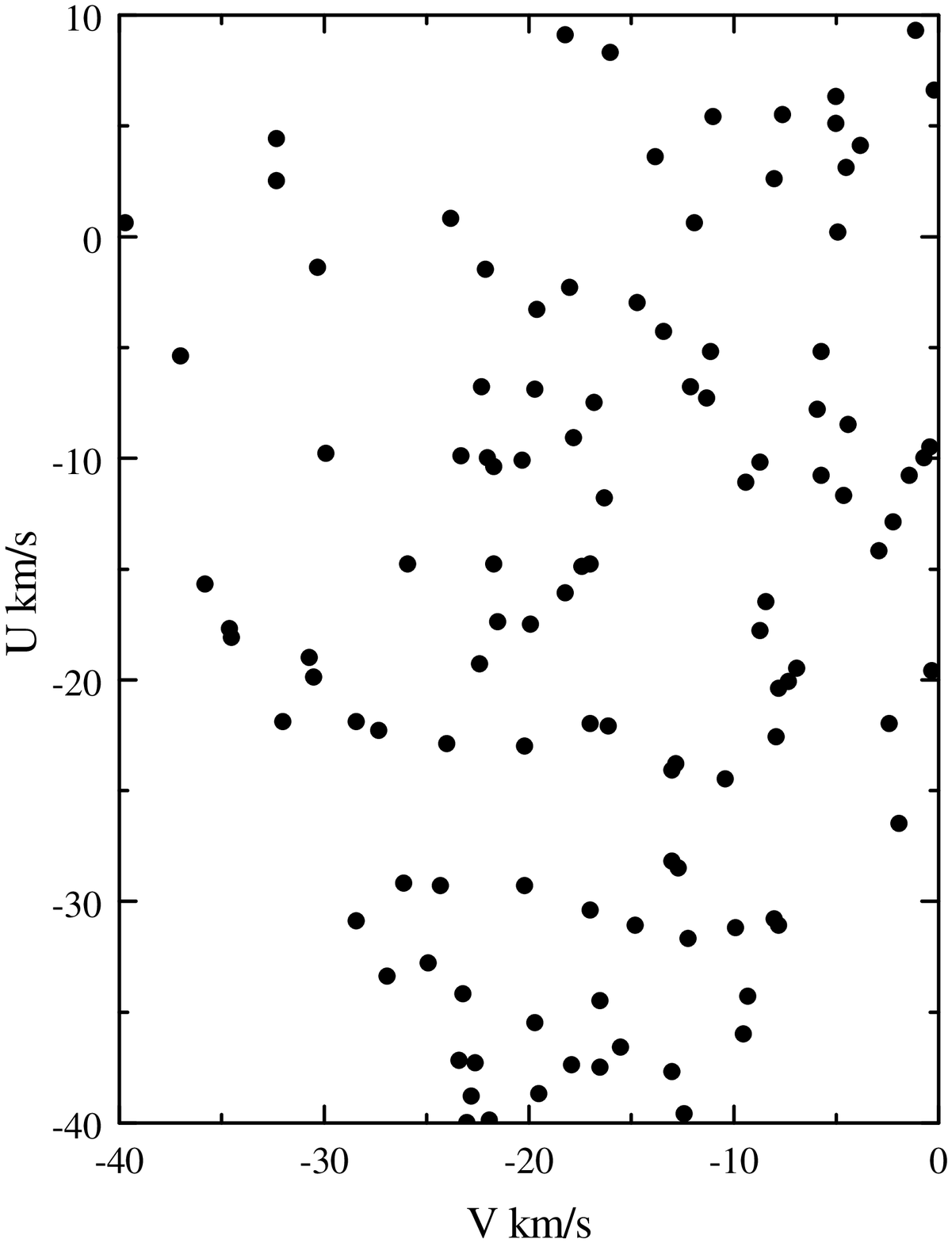] {(U, V) space velocity components for the Hipparcos stars  with $\pi>0.11$ within the ER\,Eri region, excluding those 
that are possible or probable members of the HorA 
(filled circles of the central core in Fig. 1).
The space velocity components were calculated with an adopted radial velocity of \mbox{12 km s$^{-1}$} for all stars. 
Achernar is near its position in Figure\,1. 
Many high space velocity stars are outside the figure.  
There is no concentration at \mbox{(U, V) = -10, -21} -  that characterizes the HorA - showing this it is not a general behavior of the stars in the solar neighborhood.  
 \label{f2}}

\figcaption [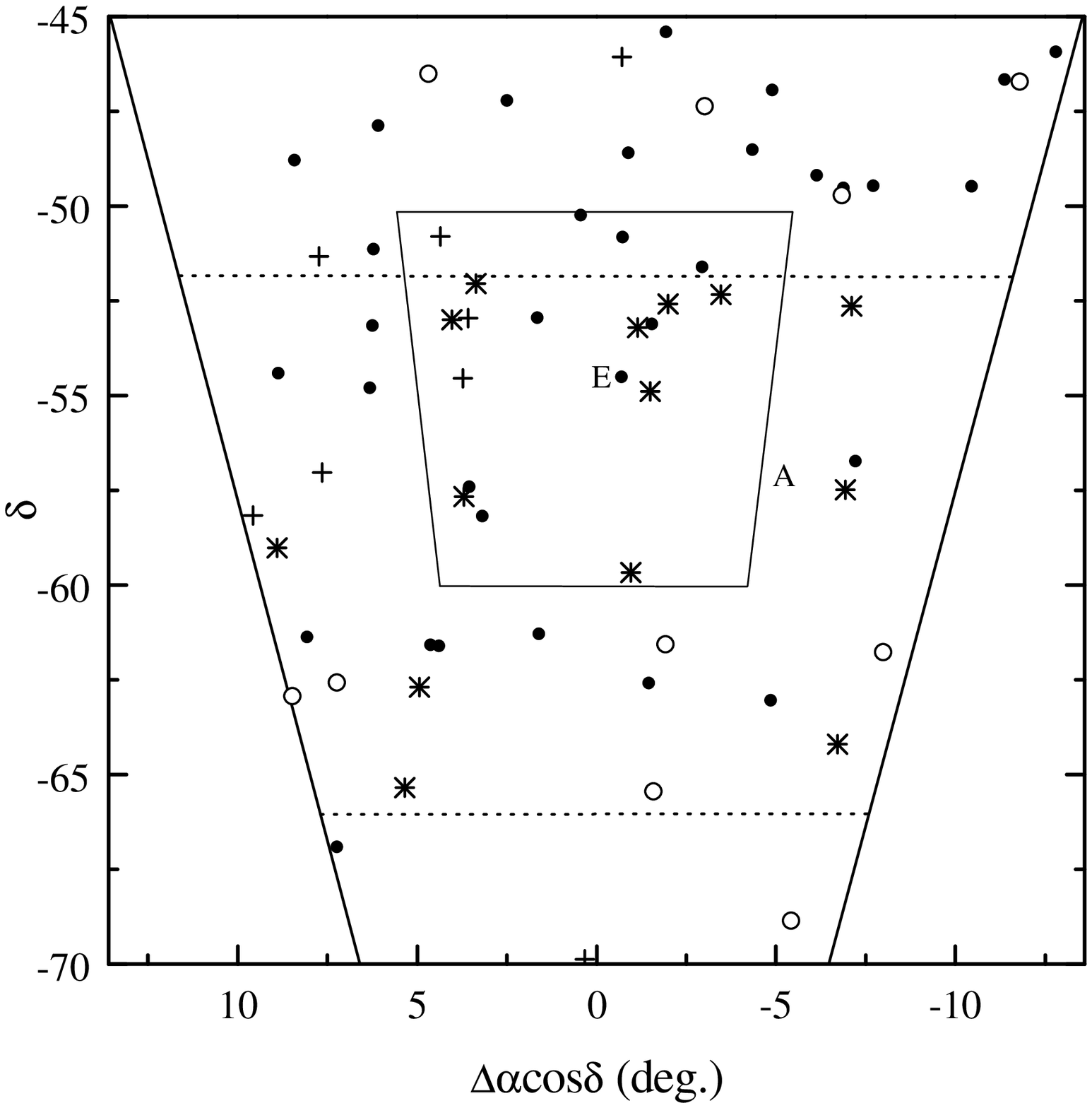] {Distribution in the sky of the selected X-ray sources in the ER Eri region. Right Ascencion is centered at 02:15. Filled circles are the field X-ray sources and open circles are the bright star counterparts of  X-ray sources not observed in this program. ER\,Eri is the filled circle marked with an E.
Asterisks represent the stars of the HorA (probable and possible members), whereas plus signs are young field stars. The position of
Achernar is marked by an A. 
The inner selected region is displayed with continuous lines and the broken lines define the {\it a posteriori} HorA region. \label{f3}}

\figcaption [Torres.fig4.ps] {Evolutive diagram for the probable  and possible members of the HorA. 
The ZAMS is indicated by a broken line. 
Filled circles -  stars with  Hipparcos parallaxes. 
Open circles - stars with kinematical parallaxes;
Open squares - ERX\,32 and ERX\,53, having no parallaxes, were plotted on the 30\,Myrs isochrone. 
The kinematical parallax of ERX\,16 seems underestimated, for its position on this diagram  and its greater distance. 
We suspect that ERX\,37S may be an undetected close binary.
We plot also the non-member young binary ERX\,33  (filled triangles). 
Most of the members have masses between 0.7 and 0.9\,{\it M}$_\sun$.
 \label{f4}}

\figcaption [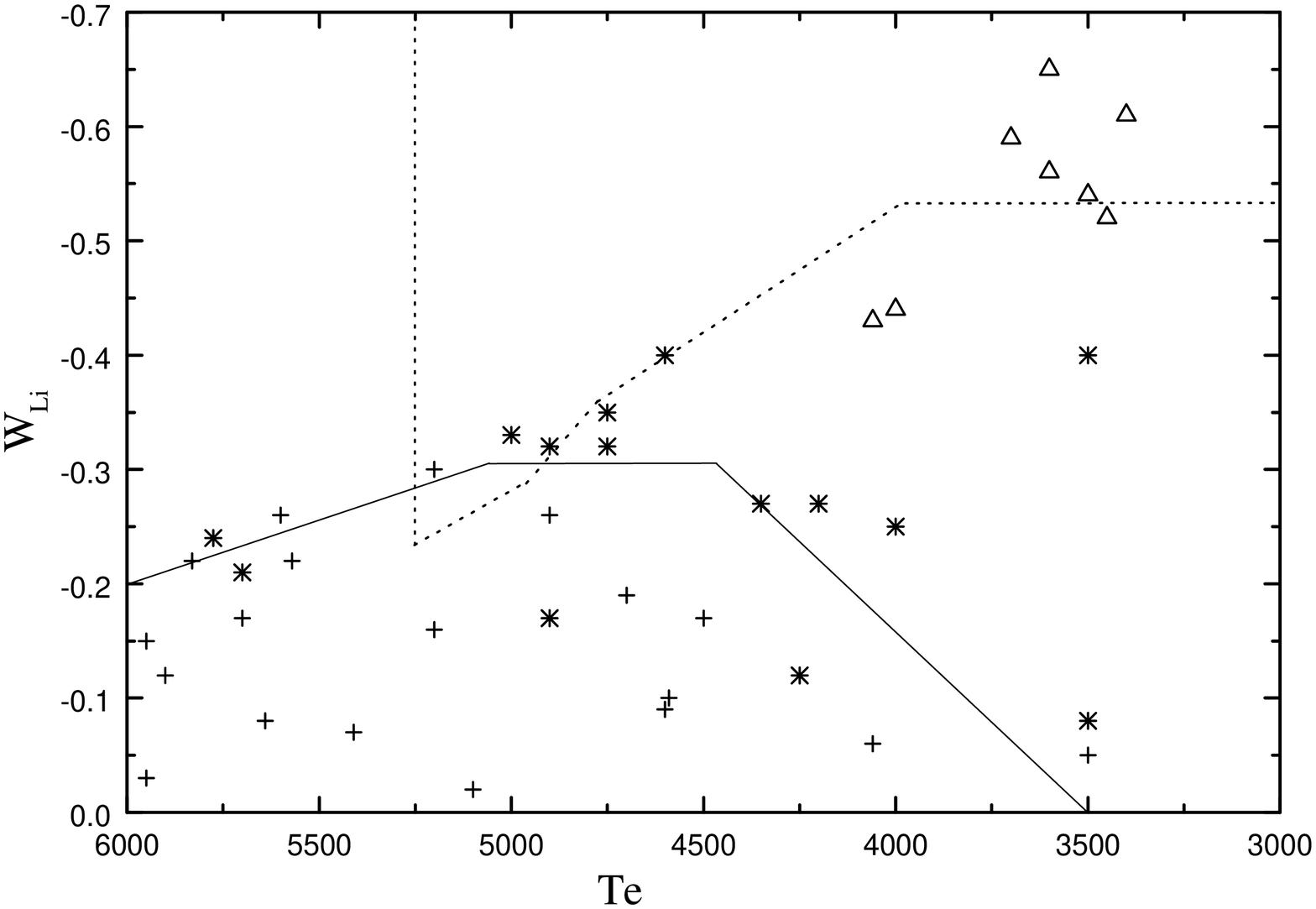] {The Li equivalent widths for the selected young stars as a function of temperature. 
Open triangles - stars of the TW Hydra Association. 
Asterisks - probable and possible members  of the Horologium Association. 
Plus signs - young field stars. 
The dashed lines, based on  \citet{28},  define the boundary for CTT. 
The solid line approximately represents the upper limit for Local Association and was taken using the plots of \citet{28} and \citet{20}. 
The HorA stars are mainly in the lower part of the post-T Tau gap, whereas the TWA stars are near the boundaries for CTT.
\label{f5}} 

\clearpage

\begin{deluxetable}{lcllrcccll}
\tabletypesize{\small}
\tablewidth {0pt}
\tablecaption {Stellar counterparts of X-ray sources in the ER\,Eri Region. \label {t1}}
\tablehead{
\colhead{ERX}&\colhead{RX J}& 
\colhead{Identification}&\colhead{$V$}&\colhead{\ub}&
\colhead{\bv}&\colhead{\vr}&\colhead{$V-I$}&\colhead{SpT}&\colhead{W$_{\it Li}$}
}

\startdata
\phn1& 010117.7$-$455632& \objectname[]{GSC\,8032-1102}&{\it 11.75} &\nodata &{\it 0.86} & \nodata& \nodata&M3Ve&\phs0:\\
\phn2 & 010835.3$-$464025& \objectname[]{HD\,6869}&\phn{\it 6.89}& {\it 0.05}&{\it 0.29} & {\it 0.17} & {\it 0.50} & F0IV & $-0.05$\\
\phn3& 011031.9$-$492928& \objectname[]{CD$\,-50\,320$}& 11.82& 0.58&0.91 &0.53 &1.05 &K2Ve& \phs0:\\
\phn4*& 011316.2$-$641142 &  \objectname[]{CPD$\,-64\,$120} & 10.29 & 0.46 & 0.86 & 0.52 & 1.01& K1Ve &$-0.330$\\
\phn5& 012221.2$-$564418 & \objectname[]{HD\,8435};\,\objectname[]{BC\,Phe}&\phn{\it 8.84} &{\it 0.32}&{\it 0.76}&{\it 0.42}&{\it 0.80}&G8Ve&\phs0:\\
\phn6*& 012320.9$-$572853 & \objectname[]{HD\,8558} &\phn8.50 & 0.22 & 0.68 & 0.40 &  0.77& G6V &$-0.205$\\
\phn7N& 012728.7$-$492822 & \objectname[]{CD$\,-50\,392$}& 11.28&0.68 &0.88 &0.54 &1.00 &G9V &\phs0:\\
\phn7S& 012728.7$-$492822 &\objectname[]{HJ\,3438B} \tablenotemark{1} &10.81&0.56 &0.84 &0.49 &0.94 &K2V&\phs0:\\
\phn8*&  012808.8$-$523824 &\objectname[]{HD\,9054};\,\objectname[]{CC\,Phe}&\phn9.07 & 0.60  & 0.91 &  0.53 &  1.01 & K2V &$-0.170$\\
\phn9& 013204.4$-$630254& \objectname[]{CPD$\,-63\,$121} &\phn9.62&0.04 &0.57 &0.35 &0.69 & F9V& $-0.04$:\\
10& 013232.8$-$493138&\objectname[]{HD\,9528};\,\objectname[]{AE\,Phe} &\phn{\it 7.78}& {\it 0.11}&{\it 0.64} &{\it 0.36} &{\it 0.71} & G2V &\phs0:\\
11&  013721.8$-$491144& \objectname[]{CD$\,-49\,451$}&10.41&1.22 & 1.42&0.93 &1.86 & M0Ve & \phs0:\\
12& 014613.2$-$465651& \objectname[]{HD\,10922} \tablenotemark{1} &\phn{\it 9.35}& {\it 0.22} &{\it 0.75} & {\it 0.43}&{\it 0.79} &G9V&\phs0:\\
13&014841.1$-$483057&\objectname[]{GSC\,8044-0859}&12.51 &0.92 &1.43 &1.01 &2.14 &M2Ve &\phs0:\\
14*&015215.7$-$521939 &\objectname[]{GSC\,8047-0232}   & 10.87 & 0.63 & 0.95  & 0.56& 1.08& K3V &$-0.350$\\
15&015557.3$-$513634&\objectname[]{HD\,11937};\,\objectname[]{$\chi$\,Eri}&\phn{\it 3.69} & {\it 0.46}&{\it 0.84} &{\it 0.46} &{\it 0.90} & G8V&\phs0:\\
16*& 020154.6$-$523453 &\objectname[]{CD$\,-53\,386$}& 11.02 & 0.61 & 0.96  & 0.58 & 1.12 &K3Ve &$-0.315$\\
17 & 020212.0$-$623532 & \objectname[]{HD\,12695}  &\phn8.10&0.04 &0.58 &0.35 & 0.69 & F9V & $-0.02$:\\
18 &  020355.4$-$452446 &\objectname[]{HD\,12759}  &\phn{\it 7.30}&{\it 0.22} &{\it 0.69} &{\it 0.39} & {\it 0.76}& G5V &\phs0:\\
19*& 020436.7$-$545320& \objectname[]{HD\,12894}&\phn6.43&$-0.04$& 0.36&0.21 & 0.43& F4V &$-0.08$:\\
20 &  020438.7$-$530722 &\objectname[]{GSC\,8483-1076} & 11.30 & 0.24 & 0.80 & 0.46&0.91 &K0Ve &\phs0:\\
21*& 020718.6$-$531155 &\objectname[]{HD\,13183}     &\phn8.63 & 0.12 & 0.65  & 0.39 & 0.76 & G5V &$-0.241$\\
22N*& 020729.1$-$594014 &\objectname[]{CD$\,-60\,416$}& 10.68 & 0.94 & 1.16 & 0.73& 1.43 & K5Ve&$-0.268$\\
22S*&020729.1$-$594014 &\objectname[]{HD\,13246} &\phn7.50 & $-0.01$& 0.52& 0.30& 0.60 & F7V & $-0.145$\\
23 &020938.1$-$483556& \objectname[]{HD\,1339}7&\phn7.79& 0.76& 1.01&0.54 &1.04&G9III &\phs0:\\
24 &  021008.2$-$543036 & \objectname[]{PDS\,1};\,\objectname[]{ER\,Eri}\tablenotemark{2}&\phn9.90 & 0.68 & 1.04   & 0.60& 1.18&K3Ve & $-0.185$\\
25 &021024.3$-$505001 & \objectname[]{HD\,13445} &\phn{\it 6.12}& {\it 0.45}&{\it 0.81} & {\it 0.42}& {\it 0.86}& G9V& \phs0:\\
26& 021055.5$-$460403 & \objectname[]{GSC\,8042-1050}   & 11.24 & 0.31  &  0.91 & 0.59  & 1.16 &K2Ve&$-0.260$\\
27& 021742.4$-$501450& \objectname[]{HD\,14366}&\phn9.29& 0.30& 0.75& 0.43& 0.83& G3V& \phs0:\\
28& 021851.4$-$695306& \objectname[]{HD\,14706}&\phn{\it 8.73}& \nodata& {\it 0.53}& \nodata&{\it 0.67}&G0V&$-0.12$\\
29& 022553.2$-$525749&\objectname[]{HD\,15279}&\phn8.37&0.11 &0.64&0.35& 0.69& G2V&$-0.05$:\\
30& 022818.9$-$611817& \objectname[]{HD\,15638} &\phn8.88&$-0.01$& 0.41& 0.26& 0.50 &{\it F3IV/V} &\nodata\\
31& 022939.1$-$471336& \objectname[]{HD\,15661}&\phn9.78& 0.38& 0.74& 0.43& 0.85&G7V&\phs0:\\
32*& 023651.8$-$520300 & \objectname[]{GSC\,8056-0482}  & 12.11 & 1.05& 1.48 & 1.08& 2.33&M3Ve&$-0.400$\\
33N& 023845.4$-$525710 & \objectname[]{HD\,16699}\tablenotemark{1}  &\phn7.86 & 0.03 & 0.52  &  0.30 & 0.58&F8V+K0&$-0.055$\\
33S&023845.4$-$525710 &\objectname[]{SAO\,232842} &\phn8.36 & 0.23 & 0.72 & 0.42 & 0.81&G7V&$-0.255$\\
34& 023901.2$-$581112&\objectname[]{CD$\,-58\,538$}&\phn9.53& 1.14& 1.44&0.93& 1.86& M0Ve&\phs0:\\
35 &024039.0$-$543252&\objectname[]{HD\,16920};\,\objectname[]{$\zeta\,$Hor}&\phn5.22& $-$0.02& 0.40&0.24 & 0.47& F4V&$ -0.065$\\
36 &024118.8$-$572522&\objectname[]{GSC\,8494-0369}&13.68&1.11&1.49&1.16&2.53&M3Ve&\phs0:\\
37N*&024146.8$-$525943 & \objectname[]{GSC\,8491-1194}  &  12.21 & 1.21 & 1.49  & 1.09& 2.40&M3Ve&$-0.08$:\\
37S*&024146.8$-$525943 & \objectname[]{CD$\,-53\,544$}& 10.21 & 1.04 & 1.26 &  0.82 & 1.60 &K6Ve&$-0.27$\\
38&024233.5$-$504756 &\objectname[]{HD\,17051};\,\objectname[]{$\iota\,$Hor}&\phn{\it 5.40}&{\it 0.08}&{\it 0.56}&{\it 0.32}&{\it 0.62}&F9V&$-0.040$\\
39*&024234.4$-$573932 &\objectname[]{GSC\,8497-0995} &  10.97 & 0.96  & 1.23  & 0.77  & 1.48 &K6Ve& $-0.120$\\
40&025111.7$-$475314&\objectname[]{GSC\,8054-0859}&11.78&0.81&1.11&0.66&1.22&K2Ve&\phs0:\\
41&025153.5$-$613704&\objectname[]{HD\,18134};\,\objectname[]{VZ\,Hor}&\phn8.84& 0.49& 0.85&0.51& 1.00&K0V&\phs0:\\
42&025346.9$-$613529&\objectname[]{GSC\,8859-0633}& 12.20& 0.84& 1.50& 1.05& 2.32&M3Ve&\phs0:\\
43&025432.4$-$510829&\objectname[]{GSC\,8057-0342}& 12.21& 1.34& 1.52& 1.07& 2.29&M3Ve&\phs0:\\
44 &025636.8$-$530934&\objectname[]{CD$\,-53\,596$}&10.17&1.07&1.10&0.68&1.26&K4V&\phs0:\\
45*& 025804.6$-$624115 &\objectname[]{GSC\,8862-0019} &  11.67 & 0.77 & 1.04 & 0.65 & 1.25 &K4Ve& $-0.395$\\
46 & 025848.7$-$544811 &\objectname[]{CD$\,-55\,479$}&10.31&1.29& 1.32&0.83& 1.59&K5V&\phs0:\\
47E &030432.9$-$511904&\objectname[]{CD$\,-51\,706$} \tablenotemark{1}&\phn8.61& 0.59& 0.86& 0.48& 0.88&K1V&$-0.020$\\
47W &030432.9$-$511904&\objectname[]{HD\,19330} &\phn7.57& 0.12& 0.56& 0.31& 0.60& G1V&$-0.030$\\
48 & 030605.0$-$484748&\objectname[]{HD\,19491} \tablenotemark{1} &\phn8.27&$-0.03$& 0.52& 0.34&0.67 &F8V&$-0.055$\\
49*&030618.8$-$652109& \objectname[]{CD$\,-65\,149$}& {\it 10.19} &\nodata & {\it 0.83}&\nodata&\nodata& K2V(e)&$-0.32$\\
50&  031113.6$-$570132 & \objectname[]{GSC\,8495-0384}  & 11.35 & 0.81  & 1.09 &  0.64 & 1.25 &K4Ve&$-0.09$\\
51& 031546.3$-$542501&\objectname[]{GSC\,8493-1147}&11.72&0.79&1.03&0.63&1.25&K3Ve&\phs0:\\
52& 032220.8$-$612253 &\objectname[]{GSC\,8860-0497}&11.39&0.92&1.12&0.74&1.45&M0Ve&\phs0:\\
53*& 032414.3$-$590101& \objectname[]{GSC\,8499-0304} &12.09 &1.13&1.25&0.78&1.54&M0Ve & $-0.245$\\
54&032738.6$-$580937& \objectname[]{CD$\,-58\,693$}&{\it 11.17}&\nodata&{\it 1.08}&\nodata&\nodata&K4Ve&$-0.100$\\
55&032837.0$-$665504&\objectname[]{GSC\,8873-0040}& {\it 10.70}& \nodata& {\it 0.73}&\nodata&\nodata& G5V(e?)&$-0.04$:\\
\cutinhead {Bright stars not observed in this program}
&010605.9$-$464309&\objectname[]{HD\,6595};\,\objectname[]{$\beta\,$Phe}&{\it 3.32}& {\it 0.57}& {\it 0.89}&{\it 0.47}&{\it 0.90}& {\it G8III}&\nodata\\
&010719.3$-$614633&\objectname[]{HD\,6793};\,\objectname[]{$\iota\,$Tuc}&{\it 5.37}& \nodata& {\it 0.88}&\nodata&{\it 0.80}& {\it G5III}&\nodata\\
&011545.7$-$685221&\objectname[]{HD\,7788};\,\objectname[]{$\kappa\,$Tuc}&{\it 4.25}& {\it 0.01}& {\it 0.48}&{\it 0.28}&{\it 0.55}& {\it F6IV}&\nodata\\
&013237.6$-$494336&\objectname[]{HD\,9544}&{\it 6.27}&\nodata & {\it 0.46}&\nodata&{\it 0.53}& {\it F4V}&\nodata\\
&015710.3$-$472307&\objectname[]{HD\,12055}&{\it 4.83}& {\it 0.52}& {\it 0.86}&\nodata&{\it 0.89}& {\it G8III}&\nodata\\
&015846.3$-$613410&\objectname[]{HD\,12311};\,\objectname[]{$\alpha\,$Hyi}&{\it 2.87}& {\it 0.14}& {\it 0.28}&{\it 0.18}&{\it 0.34}& {\it F0V}&\nodata\\
&015936.9$-$652656&\objectname[]{HD\,12452}&{\it 7.30}& \nodata& {\it 0.42}&\nodata&{\it 0.49}& {\it F3V}&\nodata\\
&024208.5$-$463121&\objectname[]{HD\,17006}&{\it 6.10}& \nodata& {\it 0.88}&\nodata&{\it 0.86}& {\it K1III}&\nodata\\
&031743.8$-$623452&\objectname[]{HD\,20766};\,\objectname[]{$\zeta^{1}\,$Ret}&{\it 5.53}& {\it 0.06}& {\it 0.64}&{\it 0.36}&{\it 0.71}& {\it G2.5V}&\nodata\\
&032920.0$-$625624&\objectname[]{HD\,22001};\,\objectname[]{$\kappa\,$Ret}&{\it 4.71}& {\it $-$0.04}& {\it 0.39}&{\it 0.24}&{\it 0.47}& {\it F5IV-V}&\nodata\\

\enddata
\tablenotetext{1}{Double line SB}
\tablenotetext{2}{Double line SB (P\,=\,5.9255 days).}
\tablecomments{Data in italic were taken from the literature (from SIMBAD or Hipparcos). 
A colon denotes less precise values.  W$_{\it Li}$  measures with three decimals come from FEROS spectra.
Asterisks mark probable and possible members of the Horologium Association.}

\end{deluxetable}

\clearpage
\begin{deluxetable}{lcllrrrrll}
\tabletypesize{\small}
\tablewidth {0pt}
\tablecaption {Stellar counterparts of X-ray sources in the Southern Region \label {t2}}
\tablehead{
\colhead{SRX}& 
\colhead{RX J}& 
\colhead{Identification}&\colhead{$V$}&\colhead{\ub}&
\colhead{\bv}&\colhead{\vr}&\colhead{$V-I$}&\colhead{SpT}&\colhead{W$_{\it Li}$}
}

\startdata
\phn1&062230.9$-$601301 &\objectname[]{HD\,45270}&\phn{\it 6.50}&{\it0.05}&{\it0.60}&{\it0.34}&{\it0.66}& G1V&$-$0.149\\
\phn2&062555.6$-$600325 &\objectname[]{GSC\,8894-0426} & {\it 12.7:}&\nodata &\nodata &\nodata &\nodata & M5Ve&\phs0:\\
\phn3&063105.2$-$590017 &\objectname[]{HD\,46697};\objectname[]{TZ\,Pic}&\phn{\it 6.63}&{\it 1.00}&{\it 1.14}&{\it 0.61}&{\it 1.13}& K1III&\phs0:\\
\phn4&063107.3$-$563914 &\objectname[]{CD$\,-56\,1546$}&{\phn\it 9.79}&\nodata &{\it 0.54} &\nodata &\nodata &G1V&\phs0:\\
\phn5&063536.4$-$623824 &\objectname[]{RR\,Pic}& {\it 12.39} &\nodata &{\it $-$0.03}&\nodata &{\it $-$0.02}& Nova&\nodata\\
\phn6&063800.7$-$613156 &\objectname[]{HD\,48189} &\phn{\it 6.19}&{\it 0.10}&{\it 0.62}&\nodata &{\it 0.69}& G0V&$-$0.150\\
\phn7&063951.7$-$612842 &\objectname[]{CD$\,-61\,1439$}&\phn{\it 9.69}&\nodata &{\it 1.38}&\nodata &{\it 1.53}& M1V&\phs0:\\
\phn8N&064207.0$-$643126 &\objectname[]{HD\,49078N}&\phn{\it 8.16}&\nodata &{\it 0.34}&\nodata &{\it 0.45}&F0V&$-$0.027\\
\phn8S&064207.0$-$643126 &\objectname[]{HD\,49078S}&{\it 10.62}&\nodata &{\it 0.72}&\nodata &\nodata &K0V&$-$0.30\\
\phn9&064345.6$-$642436 &\objectname[]{GSC\,8903-0978}&{\it 13:}&\nodata &\nodata &\nodata &\nodata &M5Ve&\phs0:\\
10&064558.4$-$550147 &\objectname[]{HD\,49595} &\phn{\it 8.76}&\nodata &{\it 0.46}&\nodata &\nodata &F5V&$-$0.05\\
11&064752.6$-$571332 &\objectname[]{GSC\,8544-1037}&{\it 11:}&\nodata &\nodata &\nodata &\nodata &K5V& $-$0.169\\
12&064812.0$-$615623 &\objectname[]{HD\,50241};\,\objectname[]{$\alpha$\,Pic}&\phn{\it 3.24}&{\it 0.12}&{\it 0.22}&{\it 0.14}&{\it 0.28}&A7V&\phs0:\\
13&064854.4$-$600857 &\objectname[]{GSC\,8895-1353} &{\it 12:} &\nodata &\nodata &\nodata &\nodata &K4Ve&\phs0:\\
14&065000.0$-$601459 &\objectname[]{HD\,50571}&\phn{\it 6.11}&\nodata & {\it 0.46}&\nodata &{\it 0.53}& F5V&\phs0:\\
15&065459.1$-$563914 &\objectname[]{CD$\,-56\,1687$}&\phn{\it 9.54}&\nodata &{\it 0.47}&\nodata &\nodata &F7V&$-$0.13\\
16&070000.6$-$612008 &\objectname[]{HD\,53143}&\phn{\it 6.81}&{\it 0.43}&{\it 0.79}&{\it 0.39}&{\it 0.82} &G9V&\phs0:\\
17&070002.4$-$570421 &\objectname[]{GSC\,8545-0020}&{\it 12:}&\nodata &\nodata &\nodata &\nodata &F7V&\phs0:\\
18&070350.6$-$582719 &\objectname[]{GSC\,8549-0141}\tablenotemark{1}&{\it 11.69}&\nodata &{\it 0.62}&\nodata &\nodata &G8V&$-$0.219\\
19&070414.7$-$625404 &\objectname[]{GSC\,8912-1753}&{\it 11:}&\nodata &\nodata &\nodata &\nodata &K2V&\phs0:\\
20&070512.3$-$573402 &\objectname[]{GSC\,8545-1235}&\phn{\it 9.87}&\nodata &{\it 0.80}&\nodata &\nodata &G9V&$-$0.070\\
21N&070922.9$-$572953 &\objectname[]{HD\,55402}&\phn{\it 9.18}&\nodata &{\it 0.44}&\nodata & &F7V&\phs0:\\
21S&070922.9$-$572953 &\objectname[]{GSC\,8558-1141}&{\it 11.08}&\nodata &{\it 0.48}&\nodata & &G2V&\phs0:\\
22&071051.1$-$573637 &\objectname[]{GSC\,8558-1148}&{\it 10.45}&\nodata &{\it 0.68}&\nodata &\nodata &G6V&$-$0.170\\
23&071059.9$-$563256 &\objectname[]{GSC\,8558-1964}&{\it 13:} &\nodata &\nodata &\nodata &\nodata & M2Ve&\phs0:\\
24&071212.8$-$611627 &\objectname[]{CD$\,-61\,1583$}&\phn{\it 9.60}&\nodata &{\it 0.48}&\nodata &\nodata & F8V&$-$0.08\\
25&071450.8$-$591600 &\objectname[]{HD\,56785}&\phn{\it 9.32}&\nodata &{\it 0.54}&\nodata &\nodata &F8V&\phs0:\\
26&071529.4$-$583228 &\objectname[]{HD\,56910}&\phn{\it 6.85}&\nodata &{\it 0.23}&\nodata &{\it 0.30}& A8V&\phs0:\\
27S&071827.7$-$572101 &\objectname[]{HD\,57555}&\phn{\it 7.90}&\nodata &{\it 0.66}&\nodata &{\it 0.72}&G5V&$-$0.05\\
27N&071827.7$-$572101 &\objectname[]{Rst\,244B}&{\it 13.6:}&\nodata &\nodata &\nodata &\nodata &F0V&$-$0.03\\
28&072123.9$-$572034 &\objectname[]{GSC\,8559-1016}&{\it 10.73}&\nodata &{\it 0.64}&\nodata &\nodata &K0V&$-$0.16\\
29&072700.5$-$561439 &\objectname[]{HD\,59487}&\phn{\it 9.18}&\nodata &{\it 0.48}&\nodata &{\it 0.55}&F7V&$-$0.08\\
30&073028.9$-$563525 &\objectname[]{GSC\,8559-1282}&{\it 11.48}&\nodata &{\it 0.71}&\nodata &\nodata &F5V&\phs0:\\
31&073113.7$-$623051 &\objectname[]{CPD$\,-62\,837$} & \phn{\it 9.77}&\nodata &{\it 0.68}&\nodata &\nodata &G8V&\phs0:\\
32&073233.9$-$562630 &\objectname[]{GSC\,8559-1292}&{\it 10.76}&\nodata &{\it 0.75}&\nodata &\nodata &G6V(e?)&\phs0:\\
33N&073341.0$-$574555 &\objectname[]{CD$\,-57\,1785$A}&{\it 10.13}&\nodata &{\it 0.36}&\nodata &\nodata & F6V&$-$0.10\\
33S&073341.0$-$574555 &\objectname[]{CD$\,-57\,1785$B}&{\it 11:}&\nodata &\nodata &\nodata &\nodata &F4V&$-$0.08\\

\enddata
\tablenotetext{1}{Double line SB}
\tablecomments{Data in italic were taken from the literature (from SIMBAD or Hipparcos). 
A colon denotes less precise values.  W$_{\it Li}$  measures with three decimals come from FEROS spectra.}

\end{deluxetable}

\clearpage

\begin{deluxetable}{lcllrrrrll}
\tabletypesize{\small}
\tablewidth {0pt}
\tablecaption {Stellar counterparts of X-ray sources in the Northern Region \label {t3}}
\tablehead{
\colhead{NRX}& 
\colhead{RX J}& 
\colhead{Identification}&\colhead{$V$}&\colhead{\ub}&
\colhead{\bv}&\colhead{\vr}&\colhead{$V-I$}&\colhead{SpT}&\colhead{W$_{\it Li}$}
}

\startdata
\phn1&094414.2$-$115218 & \objectname[]{HD\,84323}&\phn9.29&0.14& 0.66 & 0.38& 0.74 &G3V&$-$0.223\\
\phn2&094454.2$-$122047 & \objectname[]{G\,161-71}&13.80& 1.26& 1.67& 1.49& 3.37& M5Ve &\phs0:\\
\phn3&094458.6$-$134707 & \objectname[]{GSC\,5486-0589}& {\it 13:}&\nodata&\nodata&\nodata&\nodata& G0V  &\phs0:\\
\phn4&095013.6$-$104255 & \objectname[]{GSC\,5479-0365}&10.96 & 0.42& 0.86& 0.49& 0.94 & K0V&\phs0:\\
\phn5&095039.1$-$053029 & \objectname[]{BD$\,-04\,2739$} \tablenotemark{1} &\phn9.94&$-$0.04& 0.59& 0.36& 0.72&G0V &$-0.03$:\\
\phn6&095048.5$-$090657 & \objectname[]{G\,161-78}&13.31 &1.33& 1.66& 1.15& 2.63 &M4Ve &\phs0:\\
\phn7&095128.5$-$145041 & \objectname[]{HD\,85444}&\phn{\it 4.12} &{\it 0.65} & {\it 0.92}&\nodata&{\it 0.92}& G7III &\phs0:\\
\phn8&095351.8$-$072001 & \objectname[]{GSC\,4902-0210} &10.79& 1.19& 1.50 & 0.96& 1.91 & M1Ve&\phs0:\\
\phn9&095433.6$-$125711 & \objectname[]{HD\,85883}&\phn{\it 6.87}&\nodata&{\it 0.55}&\nodata&{\it 0.62} &F8V &$-$0.065\\
10&095509.2$-$081921 & \objectname[]{GSC\,5475-0507}&11.60& 1.19& 1.48& 0.96& 1.97 &M1Ve  &\phs0:\\
11A&095737.7$-$123003 & \objectname[]{GSC\,5484-0936A}&12.78J&1.48& 1.55 & 1.05& 2.18& M2Ve &\phs0:\\
11B&095737.7$-$123003 & \objectname[]{GSC\,5484-0936B}&12.78J&1.48& 1.55 & 1.05& 2.18& M4Ve &\phs0:\\
12&095916.0$-$073515 & \objectname[]{GSC\,5476-0149} \tablenotemark{2} &{\it 12:}&\nodata&\nodata&\nodata&\nodata&K0V&\phs0:\\
13&100035.3$-$085442 & \objectname[]{GSC\,5476-0910}&11.55&0.20& 0.78 & 0.47& 0.98   &G5Ve&\phs0:\\
14&100755.8$-$133146 & \objectname[]{HD\,87916} &\phn{\it 8.59}&\nodata&{\it 1.06}&\nodata&\nodata&G7III &\phs0:\\
15&100826.4$-$110708 & \objectname[]{HD\,87978}&\phn{\it 8.15}&\nodata&{\it 0.69}&\nodata& {\it 0.74}&G5V&\phs0:\\
16&100839.4$-$053502 & \objectname[]{GSC\,4909-0801} &{\it 13:}&\nodata&\nodata&\nodata&\nodata&K0Ve&\phs0:\\
17&101317.3$-$082702 & \objectname[]{HD\,88654} &\phn{\it 7.68}&{\it 0.53}&{\it 0.87}&{\it 0.47}&{\it 0.87}&G7V&\phs0:\\
18&101332.2$-$050512 & \objectname[]{HD\,88682}&\phn{\it 7.45}&\nodata&{\it 0.42}&\nodata&{\it 0.48}&F4V&\phs0:\\
19&101344.5$-$072301 & \objectname[]{HD\,88697} &\phn{\it 7.21}&{\it 0.03}&{\it 0.50}&\nodata&{\it 0.57}&F7V &\phs0:\\
20A&101628.3$-$052026 & \objectname[]{GSC\,4910-1132A}&{\it 13.58} &\nodata&{\it 1.47}&{\it 0.96} &\nodata&M1V(e?)&\phs0:\\
20B&101628.3$-$052026 & \objectname[]{GSC\,4910-1132B}&{\it 14.15}&\nodata&{\it $-$0.18}&{\it $-$0.05} &\nodata&{\it DAe }&\nodata\\
21&101731.3$-$080900 & \objectname[]{BD$\,-07\,3000$} &\phn{\it 9.87}&\nodata&{\it 0.71}&\nodata&\nodata&G7V& $-$0.078\\
22&101933.4$-$050610 & \objectname[]{HD\,89490}&\phn{\it 6.37}&{\it 0.60}&{\it 0.90}&\nodata& {\it 0.85} &G8V&\phs0:\\
23&101956.8$-$084150 & \objectname[]{RW\,Sex} & 10.69&$-$0.65&$-$0.01& 0.07& 0.26 &{\it DAe }&\nodata\\
\enddata

\tablenotetext{1}{Double line SB}
\tablenotetext{2}{Simbad gives a better association for this X-ray source with a QSO}
\tablecomments{Data in italic were taken from the literature (from SIMBAD or Hipparcos). 
A colon denotes less precise values.  W$_{\it Li}$  measures with three decimals come from FEROS spectra.}

\end{deluxetable}

\clearpage

\begin{deluxetable}{lcllrrrrll}
\tabletypesize{\small}
\tablewidth {0pt}
\tablecaption {Stellar counterparts of X-ray sources in the TW\,Hya Region \label {t4}}
\tablehead{
\colhead{TRX}&
\colhead{RX J}& 
\colhead{Identification}&\colhead{$V$}&\colhead{\ub}&
\colhead{\bv}&\colhead{\vr}&\colhead{$V-I$}&\colhead{SpT}&\colhead{W$_{\it Li}$}
}

\startdata

\phn1&104728.2$-$372159 & \objectname[]{GSC\,7198-0901} &   12.47& 1.01& 1.16& 0.71& 1.40& K5Ve &$-$0.029\\
\phn2&104809.7$-$320106 & \objectname[]{CD$\,-31\,8535$}&\phn9.46& 0.04& 0.60& 0.37& 0.73& G2V: &\phs0:\\
\phn3&105057.2$-$284954 & \objectname[]{CD$\,-28\,8475$}&   10.63& 0.06& 0.60& 0.35& 0.69& GOV &\phs0:\\
\phn4&105604.3$-$265309 & \objectname[]{GSC\,6643-1097} &   10.54& 0.24& 0.69& 0.38& 0.73& G5V&\phs0:\\
\phn5&105632.3$-$343404 & \objectname[]{HD\,94853};\objectname[]{OR\,Hya}&\phn8.40&0.14&0.64&0.35& 0.68& G1V&\phs0:\\
\phn6&105813.0$-$292611 & \objectname[]{GSC\,6647-0463}&     10.94& 0.26& 0.77&0.48& 0.96& K2Ve&\phs0:\\
\phn7&105841.3$-$290813 & \objectname[]{HD\,95135}     &\phn8.26 & 0.58 & 0.90&0.53& 1.03&K1III&\phs0:\\
\phn8&110110.9$-$313250 & \objectname[]{CD$\,-30\,8896$} &10.34& 0.56& 0.93 & 0.53& 1.04 & K3Ve&\phs0:\\
\phn9&110152.0$-$344212 & \objectname[]{TW\,Hya}&       11.07& -0.49& 0.97&1.01& 1.70 &K8Ve &$-$0.440\\
10N&110327.5$-$324527 & \objectname[]{GSC\,7204-0998}&13.84& 0.92& 1.52& 0.97& 2.18 & M2Ve&$-$0.05:\\
10S&110327.5$-$324527 & \objectname[]{GSC\,7204-1026}&11.62& 0.94& 1.22& 0.77& 1.53 & K7Ve&$-$0.055\\
11&110913.5$-$300133 & \objectname[]{CD$\,-29\,8887$};\objectname[]{PDS\,45}&11.07&1.13& 1.48& 1.06& 2.24 & M2Ve&$-$0.560\\
12&111028.9$-$373204 & \objectname[]{Hen\,600};\objectname[]{PDS\,50}&12.04J&0.74&1.49&1.35& 2.94& M4Ve&$-$0.610\\
13&111502.0$-$331905 & \objectname[]{HD\,97840} &\phn{\it 7.02}&\nodata&{\it 0.37}&\nodata& {\it 0.44} &F2V&\phs0:\\
14&111737.6$-$344420 & \objectname[]{HD\,98221} &\phn{\it 6.48}&\nodata&{\it 0.41}&\nodata& {\it 0.47} &F3V&\phs0:\\
15&111743.6$-$363210 & \objectname[]{HD\,98233} &\phn{\it 6.67}&\nodata&{\it 0.95}&\nodata& {\it 0.96} &G8III&\phs0:\\
16N&112117.1$-$344644 & \objectname[]{CD$\,-34\,7390$N}&10.93J&1.00& 1.44& 0.95& 1.97&M2Ve&$-$0.650\\
16S&112117.1$-$344644 & \objectname[]{CD$\,-34\,7390$S}&10.93J&1.00& 1.44& 0.95& 1.97&M1Ve&$-$0.590\\
17&112149.4$-$241116 & \objectname[]{HD\,98764} &\phn8.19& 0.18& 0.66& 0.37& 0.71&G2V &$-$0.120\\
18&112205.4$-$244632 & \objectname[]{HD\,98800};\objectname[]{TV Crt} &\phn8.90& 1.09& 1.24& 0.77& 1.51&K7V &$-$0.430\\
19&112256.2$-$254538 & \objectname[]{GSC\,6654-0317} &11.35& 0.92& 1.14& 0.66& 1.21 & K4V&\phs0:\\
20&113155.7$-$343632 & \objectname[]{CD$\,-33\,7795$};\objectname[]{PDS\,55}\tablenotemark{1}&11.37&0.93& 1.47& 1.10& 2.36 & M3Ve&$-$0.540\\
21&113241.7$-$265155 & \objectname[]{GSC\,6659-1080};\objectname[]{TWA\,8}&12.23&0.85& 1.46& 1.08& 2.41&M2Ve &$-$0.520\\
22&113259.7$-$315127 & \objectname[]{HD\,100407};\objectname[]{$\xi$\,Hya}&\phn{\it 3.65}&{\it 0.70}&{\it 0.95}&\nodata&{\it 0.92}&G8III&\phs0:\\
23&114247.8$-$354832 & \objectname[]{HD\,101799}; \objectname[]{V572\,Cen} &\phn9.32& 0.04& 0.57& 0.35& 0.68 & F8V&\phs0:\\
\enddata

\tablenotetext{1}{SB, discrepant radial velocities measured in PDS and now.}
\tablecomments{Data in italic were taken from the literature (from SIMBAD or Hipparcos). 
A colon denotes less precise values.  W$_{\it Li}$  measures with three decimals come from FEROS spectra.}

\end{deluxetable}

\clearpage

\begin{deluxetable}{lrrllrrrlcclll}

\tabletypesize{\footnotesize}
%\rotate

\tablewidth {0pt}
\tablecaption {Properties of the young stars of the ER\,Eri Region \label {t5}}
\tablehead{
\colhead{ERX}&\colhead{$\mu_\alpha$}&\colhead{$\mu_\delta$}&
\colhead{RV}&\colhead{$\pi$}&
\colhead{U}&\colhead{V}&\colhead{W}&
\colhead{M$_V$}&\colhead{$[F_x/F_b]$}&\colhead{T$_{\it eff}$}&
\colhead{W$_  {\it H\alpha}$}&\colhead{N$_{Li}$}&\colhead{vsin$\it i$}
}

\startdata
\multicolumn {14}{c} {Probable members of the HorA} \\ [0.8ex]
\hline
\vspace {-2.8 mm}
\\

\phn6&  $+93.0$ &$-36.9$ &\phn$+$9.2 &20.3   &$-10.4$ &$-22.8$ &$-0.9$&5.04 &$-3.39$&5700&$-3.3$&\phs3.2&\phn15\\
\phn8& $+104.8$ &$-43.4$&\phn$+$7.7&26.9     &$-8.8$ & $-19.5$ &$-1.0$&6.22 &$-3.31$&4900&$-0.2$&\phs2.1&\phn\phn6\\
14&  $+47.0$ &$-7.5$&$+$13.8 &{\it 11.2}     &$-11.3$ &$-20.7$ &$-6.3$&6.12 &$-3.08$&4750&$-0.2$&\phs3.2&\phn19\\
16&  $+35.0$ &$-9.0$&$+$11.9&\phn{\it 8.5}   &$-9.6$ & $-21.1$ &$-3.2$&5.67 &$-3.08$&4750&$+0.1$&\phs2.9&\phn19\\
21& $+85.2$ &$-22.5$&\phn$+$9.9&19.9         &$-9.8$ & $-21.0$ &$-0.6$&5.12 &$-3.24$&5775&$-3.7$&\phs3.3&\phn23\\
22N& $+85.9$&$-24.0$& $+$10.0 &22.2          &$-7.9$ & $-20.0$ &$-0.3$&7.41 &${\it -3.22}$&4350&$+0.7$&\phs1.8&\phn12\\
37S& $+101.6$&$-9.9$&\phn$+$9& {\it 25.0}    &$-10.4$ &$-18.6$ &$1.1$&7.20 &$-2.97$ &4200&$+1.4$&\phs1.9&\phn80\\
39& $+82.2$&$-8.9$& $+$12.4 & {\it 20.8}     &$-9.1$ & $-20.6$ &$-1.3$&7.56 &$-3.52$ &4250&$+0.4$&\phs0.9&\phn\phn5\\
45& $+39.4$&$-0.4$& $+$16.3& {\it 10.2}      &$-8.3$ & $-22.8$ &$-3.6$&6.71 &$-3.13$ &4600&$+0.2$&\phs3.3&\phn\phn6 \\
49&$+44.7$&$+7.1$&$+$14&{\it 11.9}           &$-9.9$  &$-20.4$ &$-2.6$&5.57&$-3.02$&4900&$-0.3$&\phs3.1&\phn75\\
\cutinhead {Possible members of the HorA}

\phn4&  $+72.0$ &$-6.0$&\phn$+$6.3&{\it 17.1}&$-13.5$ &$-15.9$& $-2.3$& 6.47&$-2.91$ & 5000& $+0.3$&\phs3.4&\phn32\\
19&  $+75.8$ &$-24.7$& $+$13.3 &21.2         &$-6.9$ & $-20.7$ &$-4.3$&3.06 &$-4.94$&6600&$-7.0$&\phs3.0:&110\\
22S& $+93.0$&$-18.8$& $+$11& 22.2            &$-9.6$ & $-20.9$ &$-1.2$&4.23 &${\it -4.26}$&6280&$-4.4$&\phs3.3&\phn35\\
32&\nodata&\nodata&$+$16& {\it 24}           &\nodata&\nodata &\nodata&9.01&$-2.90$ &3500& $+5.0$ &\phs1.3 &\phn37\\
37N&$+97.5$&$-13.7$&$+$14&{\it 25.0}         &$-9.2$&  $-21.2$& $-3.2$&9.20&$-3.22$&3500&$+5.8$& $-1.2$:&\phn\phn8:\\
53&\nodata&\nodata&$+$16.6& {\it 13}         &\nodata&\nodata &\nodata&7.66 &  $-3.42$&4000&$+0.8$&\phs1.0&\phn11\\
\cutinhead { Stars rejected as members of the HorA}

24\tablenotemark{1}&$+39.8$&$+24.6$&\phn$+$9.0&\phn {\it 3}&$-70.8$&$-22.6$&$-4.9$&2.29&$-2.90$&4700&$+0.6$&\phs1.7&\phn38\\
26*&  $+56.1$&$-22.2$ &$+$25.4& {\it 12}       &$-9.6$ & $-30.5$ & $-13.9$&6.64 &$-3.00$ &4900& $+0.2$&\phs2.6&\phn36\\
28& $+1.6$&$+1.2$&\phn$+$4&14.8               &$+0.5$&$-2.8$&$-2.9$&4.58&$-4.13$&6030&$-3.5$&\phs3.0&\nodata\\
33N& $+72.7$&$+48.8$&$+$15.9&16.3             &$-24.0$ & $-15.1$ & $-9.9$&3.92 &${\it -3.75}$&6200&$-5.6$&\phs2.5&\phn22\\
33S& $+72.7$&$+48.8$&$+$15.9&16.3             &$-24.0$ & $-15.1$ & $-9.9$&4.42 &${\it -3.58}$&5600&$-3.5$&\phs3.1&\phn\phn8\\
35&  $+32.8$&$+5.2$&\phn$+$5.8& 20.5 &         $-5.4$ & $-7.6$&$-2.2$&{\it 2.53}&$-5.29$&6600&$-6.6$&\phs2.9&\phn\phn8\\
38*& $+333.7$&$+219.2$& $+$16.6 & 58.0         &$-31.2$&$-17.2$&$-8.5$&4.22&$-4.98$&6125&$-4.3$&\phs2.4&\phn\phn7\\
47E*& $+85.4$&$+72.2$& $+$20.0 & 18.2          &$-28.8$&$-17.2$&$-11.0$&4.91&${\it -4.76}$&5100&$-2.5$&\phs1.0:&\phn\phn6\\
47W*& $+88.2$&$+71.9$&$+$20.1 & 18.6           &$-28.6$&$-17.6$&$-10.9$&3.92&${\it -5.10}$&5950&$-4.5$&\phs1.6:&\phn\phn6\\
50& $+2.6$&$+24.9$&$+$57& {\it 16}            &$-4.7$ & $-33.2$ & $-46.7$&7.37 &$-3.21$ &4600&$+1.0$&\phs1.2&\phn35\\
54 &$+78.6$&$+74.8$&$+$81&{\it 18}            &$-24.8$&$-60.2$&$-56.0$&7.45&$-2.67$&4590&$+2.0$&\phs1.3&\phn35\\

\enddata

\tablenotetext{1}{This star, \objectname[]{ER\,Eri}, is a RS CVn-like system and, therefore, not young.}
\tablecomments{Data in italic are evolutive or kinematical paralaxes, as explained in the text, or deduced values, for \mbox{M$_V$} and \mbox{$[F_x/F_b]$}, taking into account duplicity.
A colon denotes less precise values. Asterisks mark stars in the northern strip.}

\end{deluxetable}

\clearpage
\begin{deluxetable}{lrrllrrrrccrcr}
                   
\tabletypesize{\footnotesize}
%\rotate

\tablewidth {0pt}
\tablecaption {Properties of the young stars of the control regions \label {t6}}
\tablehead{
\colhead{$\ast$}&\colhead{$\mu_\alpha$}&\colhead{$\mu_\delta$}&
\colhead{RV}&\colhead{$\pi$}&
\colhead{U}&\colhead{V}&\colhead{W}&
\colhead{M$_V$}&\colhead{$[F_x/F_b]$}&\colhead{T$_{\it eff}$}&
\colhead{W$_  {\it H\alpha}$}&\colhead{N$_{Li}$}&\colhead{vsin$\it i$}
}

\startdata

\multicolumn {14}{c} {Southern Region} \\ [0.8ex]
\hline
\vspace {-2.8 mm}
\\
\phn1&  $-11.2$  & $+64.2$ &$+$32 & 42.6&$-7.5$ & $-28.4$ & $-14.6$& 4.65&$-4.16$ & 5950& $-4.0$&3.1&18\\
\phn6&  $-50.1$ &$+72.7$ &$+$35.0 &46.2 &$-7.7$ & $-30.3$ & $-18.1$&4.51 &$-3.71$&6030&$-3.5$&3.2&18\\
\phn8S& $-56.0$&$-21.0$&$+${\it 30:} &{\it 10}&$+6\phd\phn $&$-15$\phd\phn &$-38$\phd\phn &5.62 &$-4.28$&5200&$-0.4$&3.5&\nodata\\
11&  $+2.2$ &$+18.5$&$+$30.5 &{\it 20} &$-5.6$ & $-28.5$ & $-10.4$&7.5:\phd &$-3.72$&4500&$-0.6$&1.4&7\\
18&$+8.0$&$+31.1$&$+${\it 30:} \tablenotemark{1}&\phn{\it 4}&$-33$\phd\phn&$-35$\phd\phn&$+8$\phd\phn&4.70&$-3.38$&5570&$-2.7$&3.2&60\\
20&  $+2.4$ &$-21.1$& $-$45.2 & {\it 17}&$+7.3$ & $+42.4$ & $14.9$&6.02&$-3.74$&5410&$-2.1$&1.9&25\\
22&$-10.7$&$+8.4$&$+$29.7&\phn{\it 9}   &$-6.9$&$-26.4$&$-13.8$&5.22 &$-3.28$&5700&$-2.8$&3.0&27\\
28& $-1.6$&$+22.1$& $+${\it 30:}&{\it 11}&$-10$\phd\phn & $-29$\phd\phn & $-7$\phd\phn&5.94 &$-3.05$&5200&$-2.0$&2.5&\nodata\\
\cutinhead {Northern Region}
\phn1&$-42.2$&$-30.7$& $+$19.6&12.5&$-12.1$&$-24.0$&$-7.2$&4.77&$-3.32$ &5830& $-3.3$ &3.5 &30\\
21& $-23.0$&$+3.2$&$+$49.8&{\it 13}&$-20.2$&$-37.5$&$+27.1$&5.44 &$-3.48$&5640&$-1.8$&2.2&9\\
\cutinhead {TW\,Hya Region}

\phn9&  $-66.9$&$-12.4$&$+$13.0 & 17.7& $-12.0$ & $-18.2$&$-5.0$&7.31&$-2.52$&4000&$+300$\phd\phn&2.5&14\\
10N&$+${\it 31.7}&$-${\it 33.2}&\phn$-${\it 7.6}&{\it 20}& $+$9.4&$+$6.7&$-$6.6&10.35&$-${\it 2.52}&3500&$+5$\phd\phn&-0.7:&\nodata\\
10S& $+31.7$&$-33.2$&\phn$-$7.6&{\it 20}&$+9.4$ & $+6.7$ & $-6.6$&8.12 &$-${\it 3.12}&4060&$+1.0$&0.2&3\\
11& $-$90.1&$-$21.1&$+$12.0 & {\it 23.2} &$-12.4$&$-18.0$&$-4.9$&7.89&$-3.20$&3600&$+2.1$&3.5&15\\
12& $-$112\phd\phn&$-$11\phd\phn&$+$14.0&{\it 28.7}&$-12.5$&$-19.3$&$-3.5$&{\it 10.08}&$-3.18$ &3400&$+10$\phd\phn&3.5&15\\
16N&\nodata&\nodata&$+$11.3&{\it 20}&\nodata&\nodata &\nodata&{\it 8.19}&$-2.86$&3600&$+3.5$&3.6&10 \\
16S&\nodata&\nodata&$+$11.6&{\it 20}&\nodata&\nodata &\nodata&{\it 8.19}&$-2.86$&3700&$+2.7$&3.6&12\\
17& $+40.5$&$-33.7$&\phn$+$0.3 &20.8&$+11.7$&$-0.4$&$-2.5$&4.78&$-4.37$&5900&$-3.8$&2.8&4\\
18&$-85.4$&$-33.4$&$+$13&21.4&$-11.4$&$-20.7$&$-4.8$&{\it 6.30}&$-3.42$&4060&$+0$\phd\phn&2.4&10\\
20& $-81.6$&$-29.4$&$+$14&{\it 20.9}&$-10.1$ & $-21.3$&$-5.2$&7.97 &$-2.97$ &3500&$+10$\phd\phn&3.7&58\\
21&\nodata&\nodata&\phn$+$7.5&{\it 20}&\nodata&\nodata&\nodata&8.74&  $-2.69$&3450&$+25\phd\phn$&3.3&5\\
\enddata
\tablenotetext{1}{Double line SB, velocity adopted as explained in the text.}
\tablecomments{Data in italic: for RV, we adopted  a typical value of the measured stars in the Southern Region or, for TWX10, the  one of the brigth component; for parallaxes, estimated or kinematical values, as explained in the text; for \mbox{M$_V$} and \mbox{$[F_x/F_b]$},  deduced values taking into account the duplicity. A colon denotes less precise values.}

\end{deluxetable}
\clearpage

\begin{deluxetable}{llrrlc}
\tablewidth {0pt}
\tablecaption { Non-X Ray stars possibly belonging to the HorA  \label {t7}}
\tablehead{ 
\colhead{Identification}&\colhead{$V$}&\colhead{\bv}&
\colhead{$V-I$}&\colhead{SpT}&\colhead{$\pi$}
}
\startdata

 \objectname[]{HD\,8077}   & 8.88& 0.55& 0.62& F6V&11.6\\
 \objectname[]{HD\,10144}  & 0.45& $-$0.16& $-$0.17&  B3Vpe&22.7\\
 \objectname[]{HD\,10269}  & 7.08& 0.47& 0.54& F5V&20.4\\
 \objectname[]{HD\,10472}  & 7.62& 0.42& 0.49& F2IV/V&15.0\\
 \objectname[]{HD\,14228}  & 3.56& $-$0.12& $-$0.11& B8IV-V&21.1\\
 \objectname[]{HD\,20888}  & 6.03& 0.13& 0.15& A3V&17.2\\
\tablecomments{Data from Hipparcos.} 
 
\enddata

\end{deluxetable}
\clearpage

\begin{deluxetable}{cccc|ccc|cc}
\tablewidth {0pt}
\tablecaption {Summary of the statistics of the observed regions \label {t8}}
\tablehead{
\colhead{Region}&\colhead{Area}&
\colhead{X-ray}& 
\colhead{D$_1$}&\colhead{XY}&\colhead{Fraction}&
\colhead{D$_2$}&\colhead{XA}&\colhead{D$_3$}
}

\startdata

I-ERX&115 & 19 &  0.17&  8& 0.42& 0.07&  7&0.06\\
O-ERX&388 & 46 &  0.12& 11& 0.24& 0.03&  6&0.02\\
ERX&503& 65 &     0.13& 19& 0.29& 0.04& 13&0.03\\
NERX&174& 24&     0.14&  2& 0.08& 0.01& \nodata&\nodata\\
HRX&270& 38 &     0.14& 16& 0.42& 0.06& 13&0.05\\
SRX&100& 33 &     0.33&  8& 0.24& 0.08& \nodata&\nodata\\
NRX&100& 23 &     0.23&  2& 0.09& 0.02& \nodata&\nodata\\
TRX& 182& 23&     0.13&  9& 0.39& 0.05& 7&0.05\\

\tablecomments{The first two regions are the inner and outer ER\,Eri regions, as defined in section\,2. 
NERX and HRX are the northern strip and the HorA region as defined in section\,5. 
In column 3 the number of all selected X-ray sources with stellar counterparts is given, as explained in section\,2; 
D$_1$ is their sky density in each area (number of stars per square degrees). 
XY means the number of the X-ray sources associated with possible young stars later than F;
in column 6 we present their fraction relative to the selected X-ray sources; 
D$_2$ is their sky density in each area. 
XA is the number of X-ray sources associated with probable and possible young stars later than F belonging to the HorA or to the TWA  in the observed regions,  and D$_3$ is their density.}

\enddata

\end{deluxetable}

\clearpage

\begin{deluxetable}{lclllrrr}
\tablewidth {0pt}
\tablecaption {Photometric data of other possible members of the TWA \label {t9}}
\tablehead{
\colhead{name}&
\colhead{RX J}& 
\colhead{Identification}&\colhead{$V$}&\colhead{\ub}&
\colhead{\bv}&\colhead{\vr}&\colhead{$V-I$}
}

\startdata

\objectname[]{TWA\,6}&101828.7$-$315002 & \objectname[]{GSC\,7183-1477} &   11.62& 1.02& 1.31& 0.84& 1.68\\
\objectname[]{TWA\,7}&104230.3$-$334014 & \objectname[]{GSC\,7190-2111} &   11.65& 0.98& 1.46& 1.09& 2.44\\
\objectname[]{TWA\,9A}&114824.5$-$372838 & \objectname[]{CD$\,-36\,7429A$}&  11.26& 0.97& 1.26& 0.80& 1.62\\
\objectname[]{TWA\,9B}&114824.5$-$372838 & \objectname[]{CD$\,-36\,7429B$}&  14.00& 1.4:& 1.43& 1.17& 2.58\\
\objectname[]{TWA\,10}&123504.4$-$413629 & \objectname[]{GSC\,7766-0743}&    12.96&0.81&1.43&1.09& 2.47\\
Sterzik\,1 \tablenotemark{1}&110940.0$-$390657& \objectname[]{CD$\,-38\,6968$}&\phn9.92& 0.07& 0.59&0.34& 0.67\\
Sterzik\,2&(1121.1$-$3845)& \objectname[]{GSC\,7739-2180}     & 12.85 & 1.0: & 1.53&1.08& 2.36\\
\tablenotetext{1}{Not a member}
\tablecomments{These objects come from \citet{39} and \citet{36}. A colon denotes less precise values.}

\enddata

\end{deluxetable}

\end{document}